\documentclass[10pt]{article}
\usepackage{geometry}                
\usepackage{makeidx}
\usepackage[pdftex]{graphicx}
\usepackage{lscape}
\usepackage{amssymb}
\usepackage{epstopdf}
\usepackage{amsmath,amssymb,amsthm}
\usepackage{latexsym}
\usepackage{listings}
\usepackage{chemarr}
\pdfoutput=1

\newtheorem{axiom}{Axiom}[section]
\newtheorem{defin}{Definition}[section]
\newtheorem{theo}{Theorem}[section]
\newtheorem{propo}{Proposition}[section]
\newtheorem{lemma}{Lemma}[section]
\newtheorem{corol}{Corollary}[section]
\newtheorem{remark}{Remark}[section]
\newtheorem{problem}{Problem}[section]
\newcommand{\bax}{\begin{axiom}}
\newcommand{\eax}{\end{axiom}}
\newcommand{\bass}{\begin{assumption}}
\newcommand{\eass}{\end{assumption}}
\newcommand{\bdefi}{\begin{defin}}
\newcommand{\edefi}{\end{defin}}
\newcommand{\bth}{\begin{theo}}
\renewcommand{\eth}{\end{theo}}
\newcommand{\bprop}{\begin{propo}}
\newcommand{\eprop}{\end{propo}}
\newcommand{\blem}{\begin{lemma}}
\newcommand{\elem}{\end{lemma}}
\newcommand{\bcor}{\begin{corol}}
\newcommand{\ecor}{\end{corol}}
\newcommand{\brem}{\begin{remark}}
\newcommand{\erem}{\end{remark}}
\newcommand{\bprob}{\begin{problem}}
\newcommand{\eprob}{\end{problem}}
\newcommand{\bpf}{\begin{proof}}
\newcommand{\epf}{\end{proof}}

\newcommand{\bx}{{\bf x}}
\newcommand{\field}[1]{\mathbb{#1}}

\newcommand{\R}{\field{R}}

\newcommand{\pa}{\partial}

\newcommand{\ra}{\rightarrow}

\newcommand{\rlha}{\xrightleftharpoons}
\newcommand{\de}{\delta}

\newcommand{\eps}{\epsilon}

\newcommand{\Lop}{{\widehat{{\mathcal L}}}}
\newcommand{\Kd}{{{\mathcal K}}}

\newcommand{\Sig}{{\Sigma}}
\newcommand{\sig}{{\sigma}}

\newcommand{\vI}{\vec{{\mathcal I}}}
\newcommand{\ve}{\vec{e}}
\newcommand{\noi}{\noindent}
\newcommand{\cX}{{\mathcal X}}
\newcommand{\rank}{\mbox{rank}}
\newcommand{\Ra}{\Rightarrow}

\begin{document}

%
%

\title{An Averaging Principle for Combined Interaction Graphs.  Connectivity and Applications to Genetic Switches.}

\author{Markus Kirkilionis\\ 
\texttt{mak@maths.warwick.ac.uk}\\
University of Warwick, Mathematics Department,\\ 
and\\
Luca Sbano\\ \texttt{luca.sbano@istruzione.it}\\
IISS Von Neumann, Roma.
}

\maketitle

\begin{abstract}
Time-continuous dynamical systems defined on graphs are often used to model complex systems with many interacting components in a non-spatial context. In the reverse sense attaching meaningful dynamics to given 'interaction diagrams' is a central bottleneck problem in many application areas, especially in cell biology where various such diagrams with different conventions describing molecular regulation are presently in use.  In most situations these diagrams can only be interpreted by the use of both discrete and continuous variables during the modelling process, corresponding to both deterministic and stochastic hybrid dynamics. The conventions in genetics are well-known, and therefore we use this field for illustration purposes. In \cite{LM1} and \cite{LM2} the authors showed that with the help of a multi-scale analysis stochastic systems with both continuous variables and finite state spaces can be approximated by dynamical systems whose leading order time evolution is given by a combination of ordinary differential equations (ODEs) and Markov chains. The leading order term in these dynamical systems is called \emph{average dynamics} and  turns out to be an adequate concept to analyse a class of simplified hybrid systems. Once the dynamics is defined the mutual interaction of both ODEs and Markov chains can be analysed through the (reverse) introduction of the so called \emph{Interaction Graph}, a concept originally invented for time-continuous dynamical systems, see \cite{Domijan}. Here we transfer this graph concept to the average dynamics, which itself is introduced as an heuristic tool to construct models of reaction or contact networks.  The graphical concepts introduced form the basis for any subsequent study of the qualitative properties of hybrid models  in terms of connectivity and (feedback) loop formation. \end{abstract}


\section{Introduction}
The modelling of systems with many interacting components usually involves employing different mathematical methods and approaches in order to obtain effective descriptions. A typical such situation is the study  of  cellular systems involving different  interactions of (bio-)molecular species establishing some cellular function. In this class of problems many processes occur simultaneously,  and they are so interconnected that the concept of a network turns out to be a natural  framework to work with. It has become standard in cell biology to draw diagrams to illustrate the molecular systems, where only part of the components and arrows can be straightforwardly interpreted as classical chemical reactions. In order to introduce a well-defined network structure it is necessary to define a set of interpretation rules relating the system components to the  network elements. The latter are given by the vertices and edges of the network (compare this with the introduction in \cite{review}). There is however no unique such mapping even if the network components are well defined. This has been for example illustrated in  \cite{Domijan} where different graphs are shown that are associated to mass-action reaction networks. But nevertheless there is a quite generic such network structure  given by the so-called  \emph{interaction graph} $\vI$ (denoted as $G_I$ in  \cite{Domijan} ). For the setting in this paper the classical interaction graph $\vI$ needs to be extended, as we are now looking at situations where some of the players or particles are not abundant enough to be described by a continuous variable. The vertex set (nodes) becomes larger, there is a node set containing components describing a single continuum variable (like a chemical concentration, which in this framework are the macro-states modeled with deterministic dynamics), and nodes associated to every discrete state of the Markov chain describing microscopic interactions. The approach taken in this article is purely graphical. i.e. we define vertex and edge sets of graphs associated to hybrid dynamics and their scalings. We emphasise that alternatively one could introduce the respective adjacency and incidence matrices and present all graph concepts introduced on an pure algebraic level, \cite{Biggs,Godsil}. Especially purely algebraic conditions for the uniqueness of an invariant measure of a Markov chain will be much more familiar to most readers, see section \ref{subsec:mc} where we use graph connectivity as a concept instead. Knowing about this mathematical duality we therefore restrict ourselves here to the graph aspect only in order to (a) emphasise the connection to (dynamical) network theory, and (b) showing the connection to interpretation of (biological) graphical diagrams in a complete and consistent way. \\
 
 In this paper we always assume spatial homogeneity. We are interested in a special class of systems which are characterised by  the presence of two different kinds of variables. One variable subset admits a description through the notion of continuity (or concentration) based on the assumption of \emph{mass action} kinetics, which itself is based implicitly by a so-called continuum limit. Therefore ordinary differential equations (ODE) are used to describe their dynamics. The other variables can only take finitely many possible states and are  inherently stochastic. Hence the overall system is typically \emph{hybrid}. Hybrid systems have been used to describe many different situations, see for instance \cite{hybrid1}. Very close to our framework are \cite{julius}, and the late work by Gillespie \cite{Gillespie}. Note that in our framework the  dynamics of both the discrete and the continuous state spaces have in principle a similar mathematical description. In fact, any set of reactions has to be described via stochastic dynamics, typically a \emph{master equation} (ME) giving the time evolution of the probability of the system while being in a certain state at time $t$.  As described in \cite{LM1,LM2,LM3, review} microscopically  the state of a reaction or 'rule-based' system is an element in ${\mathbb L}_\de\times\Sigma$, where ${\mathbb L}_\de$ is a lattice, $\de$ is its scale -- possibly a vector  as particle properties may scale differently for each species -- and $\Sigma$ is a finite set, the discrete state space. The 'continuous' state space ${\mathbb L}_\de$ accommodates the states  of the reactions or rules that can be described by mass-action kinetics in a large number limit,  whereas $\Sig$ contains the states that are modelled intrinsically discrete. Typical examples are  conformational changes and binding/unbinding events of macro-molecules, but some switch in an engineering example is leading towards the same kind of description.  In a larger system there are in general many molecular  subsystems whose states are finite and which  participate in cellular  regulation.  In a cell biology context we refer to such subsystems as   \emph{molecular machines}.  The consequence of this is that  $\Sig$ collects the possible states of all the molecular machines and hence $\Sig$ is naturally given by  the Cartesian product of the discrete state spaces of each such subsystem. In graph terms this generally means that the Markov chain transition graph has as many disconnected subgraphs (internally connected components) as there are different such molecular machines. The influence of the algebraic structure of $\Sig$ on the dynamics has been partially  studied in \cite{LM2}. In \cite{LM1} the continuum limit

 $${\mathbb L}_\de\ra\R^N\mbox{ as $\de\ra 0$}$$

 \noi was introduced. In such a regime the occupation numbers in ${\mathbb L}_\de$ become concentrations $\bx$ in $\R^N$, but the states  in $\Sig$ retain their discrete structure. The ME becomes a vectorial Fokker-Planck equation of the form:
 
  \begin{equation}
  \label{FPE}
  \frac{\pa \rho_\sig(t,\bx)}{\pa t}=\sum_{\sig'\in\Sig}\Lop^*_{\sig\sig'}[\bx]
  \cdot \rho_{\sig'}(t,\bx)+\frac{1}{\eps}\,\sum_{\sig'\in\Sig}\Kd^T_{\sig\sig'}(\bx)
  \cdot \rho_{\sig'}(t,\bx)~~\forall \sig\in\Sig.
\end{equation}

Here $\rho_\sig(t,\bx)$ is a probability density,

$$\int_{\R^N}d\bx\sum_{\sig\in\Sig}\rho_\sig(t,\bx)=1,$$

and $\eps$ is a time scale describing the ratio of speed of changes in the deterministic part -- compared to the switching of states in the discrete part of the state space described by the Markov chain. The differential operator $\Lop[\bx]$ is matrix-valued and  describes the reactions of mass-action type, whereas $\Kd(\bx)$ is a Markov chain (MC) generator which describes the interactions in $\Sig$. The parameter $\eps$ is a scaling factor involved in the continuum limit and used to derive the differential operator. Such a parameter is usually infinitesimal and hence (\ref{FPE}) becomes a perturbation problem as $\eps\ra 0$. It is not difficult to see that $\eps\ra 0$ corresponds to having a  processes in $\Sig$ which is much faster than a process driven by $\Lop[\bx]$. Therefore (\ref{FPE}) can be studied by  an expansion in $\eps$ through the so called \emph{adiabatic theory}  (see \cite{LM1,LM2,LM3}). The solution for $\eps=0$ corresponds to a system staying in the MC equilibrium state, which is an invariant probability measure. For $\eps>0$ sufficiently small the system evolves according to a vector field which is an average of all possible dynamics indexed by $\Sigma$, and taken against the invariant measure. The leading order term of the $\eps$-expansion of the solution of (\ref{FPE}) is a deterministic equation for the probability density (Liouville Equation). This corresponds to a deterministic vector field in the concentration space $\R^N$. We termed such a vector field the  \emph{average dynamics}.  This name for the leading part in the expansion derived by the adiabatic theory has been suggested by its derivation: the average dynamics is constructed out of  the invariant measure $\mu(\bx)$ of the MC together with the deterministic dynamics given for each state of the MC. Let $\mu(\bx)$ be the invariant measure of the MC and let  $f^{(\sig)}(\bx)$ be the deterministic vector field for each $\sig\in\Sig$. Then the average dynamics is given by \emph{averaging} the vector field against the measure $\mu(\bx)$ yielding 

\begin{equation}
\cX[\bx]=\sum_{\sig\in\Sig}\mu_\sig(\bx)f^{(\sig)}(\bx).
\label{av-vf}
\end{equation}

 \begin{figure}[htbp] 
      \begin{center}
       \includegraphics[scale=0.45]{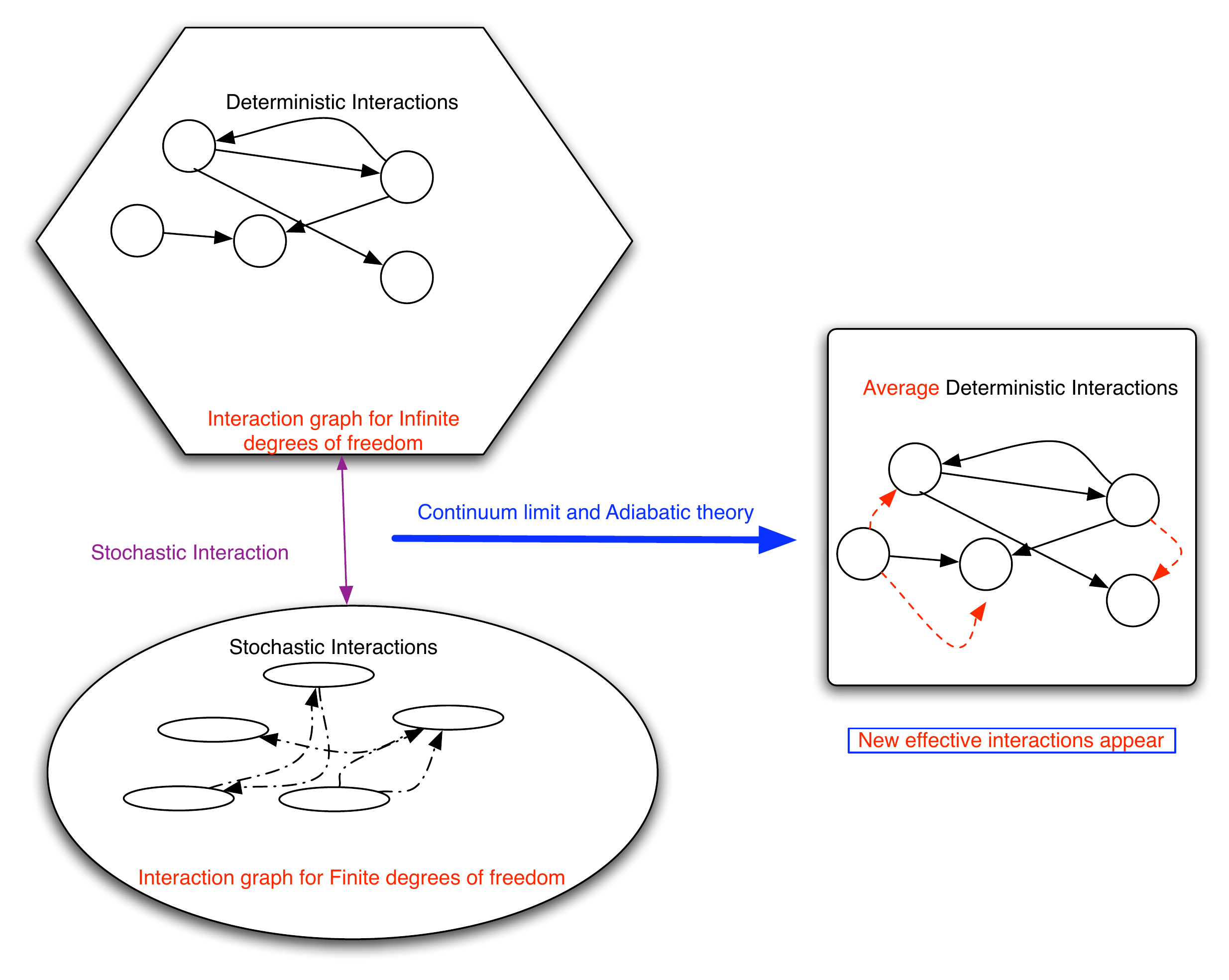} 
       \caption{A system is partitioned into two "subsystems". One subsystem is driven by deterministic interactions, which are represented by a classical interaction graph. The other subsystem has a stochastic evolution dictated by a MC that has a natural interaction graph interpretation. The two subsystem are coupled by a stochastic dynamics which is described only at the level of the  Master Equation. The continuum limit and the adiabatic theory lead to the   \emph{average dynamics}, the object is studied in this paper.}
       \end{center}
       \label{fig:system}
    \end{figure}
    
The whole scaling process can be visualized  through the diagram in Figure \ref{fig:system}.  In this paper we want to consider the averaging principle only as a heuristic tool to construct  meaningful models for applications. In this respect we present a method to derive deterministic dynamics from  \emph{hybrid} stochastic dynamics. We term this modeling procedure the \emph{averaging principle}.  We further  assume that the averaging principle applies to the specific system (which means can be modeled correctly), but also include an example where some unstructured particles do not form a density at the end of the paper. The deterministic vector field is naturally associated to an interaction graph $\vI_{D}$, and similarly another interaction graph is associated to the MC (denoted by $\vI_{MC}$). These two graphs are combined into a larger graph $\vI_{C}$ describing how stochastic and deterministic variables interact. The average dynamics describes the time evolution of the deterministic part averaged against the invariant  probability measure. In fact the invariant measure assigns probabilistic "weights"  to the finite states of the MC. To each such state corresponds a specific deterministic vector field which will give a contribution to the average vector field according to the probabilistic weight of the MC state. The new  vector field produced by the average procedure is then associated to a new (in a sense \emph{collapsed}) interaction graph $\bar{\vI}$. The new \emph{collapsed} graph resonates with the notion of \emph{condensation} of a graph \cite{WC}. In $\bar{\vI}$ the subgraph $\vI_{MC}$ is no longer present but new edges will appear. These new links connect vertices which were part of the Markov chain graph   $\vI_{MC}$ before applying the averaging procedure (see Figure \ref{fig:system}). Similar models are already known in the literature (see \cite{hybrid}), but the average dynamics  presented here is derived from basic principles that correspond (asymptotically) to the deterministic limit of the originally stochastic system. \\

The aim of this paper is to introduce and explain all the potentiality of the \emph{averaging principle} and the associated averaged interaction graph $\bar{\vI}$. We first describe the notion of the interaction graph for deterministic dynamics and for Markov chains separately, and then extend the concepts to hybrid and averaged systems.  Furthermore we investigate the consequences of the structure of  $\Sig$ from the modeling point of view.  The space $\Sig$ is in general a Cartesian product of many discrete state spaces. We show that  the structure of the infinitesimal generator $\Kd$  determines whether the discrete state (molecular) machines behave as  independent subsystems. Finally the averaging principle is illustrated by several examples taken from genetics. In particular we shall show that the dynamics of so-called \emph{feed-forward loops} as analysed in \cite{MA} can be derived more rigorously (in the sense that the underlying scalings to derive the macroscopic equations are at least known) by the average dynamics. As a final example we show that the average dynamics can be used to  re-formulate a model for genetic oscillators as introduced in \cite{VKBL}. This example shows that oscillations are produced by the interaction of two genes.  We show that  the corresponding MC can effectively often be decomposed/reduced into two independent MC's describing the dynamics of each single gene.  The evidence of the rhythms for this model is shown numerically. The modeling techniques described in this paper give also rise to model existing synthetic genomes from a dynamical perspective, with the hope to achieve that in future in a semi-automatic way once a sequence is known, see \cite{msgm-1,msgm-2}.

\section{Interaction graphs}

\subsection{The interaction graph for deterministic dynamics}
Let us consider a dynamical system defined by $N$ first-order ordinary differential equations of the form

\begin{equation}
\label{eq:det-syst}
\dot{x}_i(t)=f_i(x_1(t),\ldots,x_N(t),\alpha)+g_i(t,\beta),~~i=1,...,N.
\end{equation}

\noindent Here $\bx=(x_1,..,x_N)\in\R^N$ is the continuous state of the system, $\alpha,\beta$ are (sets of) parameters, and $f=(f_1,..,f_N)$ and $g=(g_1,..,g_N)$  are both sufficiently smooth vector fields on $\R^N$. We associate a directed graph to (\ref{eq:det-syst}) which we call the \emph{interaction graph} $\vI$ of system (\ref{eq:det-syst}).  If $g$ is not identically zero then (which implies (\ref{eq:det-syst}) is non-autonomous)  an additional state variable $x_0$ is introduced. This state $x_0$ (not affected by other states) is then called the \emph{environment}.

\bdefi[The interaction graph $\vI_{D}$ for deterministic dynamics]
The \emph{interaction graph} associated to (\ref{eq:det-syst})  is a directed graph denoted by $\vI_{D}=(V,E)$ where

\begin{itemize}

\item[(i)] the vertex set $V(\vI_{D})$ is equal to the collection $\{x_1,...,x_N\}$ of state variables in case $g \equiv 0$ in (\ref{eq:det-syst}) and $\{x_0,x_1,...,x_N\}$ otherwise,

\item[(ii)] an edge $\ve_{ij} =(x_i,x_j)$, with $i,j = 1,\ldots,N$ is an element of $E(\vI_{D})$ iff

$$\frac{\pa f_j(\bx)}{\pa x_i}\mbox{ is not identically zero.}$$ 

An edge  $\ve_{0j} =(x_0,x_j)$, with $j = 1,\ldots,N$ is an element of $E$ iff $g_i$ is not identically zero.

\item[(iii)] The edge $\ve_{ij}$ is directed from $x_i$ to $x_j$. This is equivalent to say $(x_i, x_j)$ is an ordered pair.

\end{itemize}
\edefi

\brem
The vertex $x_0$, the \emph{environment}, describes any external actions/influences on the system. Note that in applications external actions may be used to model a dynamics which is much slower than the one of the system and consequently such external variables  can be considered "frozen".  Obviously the edges of $\vI_{D}$ can be weighted with the sign of $\frac{\pa f_j(\bx)}{\pa x_i}$ telling whether an interaction is positive or negative. If such sign patterns are identical for all states $\bx$ of the respective state space, then the system is called quasi-monotone and more tools are available to characterise the qualitative properties of the system \cite{quasi}. The definition of $\vI_{D}$ can be generalised by including time dependent vector fields, that would imply a time dependence of the link structure in $\vI_{D}$ as well. But even for general autonomous ODE generating a flow the state is time-dependent, so $\vI_{D}$ does possibly change its links/edges and weights. This of course does not hold for initial conditions identical to an equilibrium.
\erem

\brem
Note that using the vector field we can readily construct the  \emph{adjacency} matrix $A$ of  $\vI_{D}$ by defining:

\[A_{ij}(\bx)=\left\{\begin{array}{ll}
1 \mbox{ if $\frac{\pa f_j(\bx)}{\pa x_i}\neq 0$}\\[3mm]
0\mbox{ otherwise}
\end{array}\right.\]

\erem

\subsection{The interaction graph for Markov chains}

A linear Markov chain (MC) has a natural structure of a directed graph which we denote by $\vI_{MC}$. 

\bdefi[The interaction graph $\vI_{MC}$ for Markov chain dynamics]
\begin{itemize}

\item[(i)] $V(\vI_{MC}) = \{\sig_1, \ldots, \sig_M \}$, where $\Sig= \{\sig_1, \ldots, \sig_M \}$ is the collection of (finite) MC states,

\item[(ii)] for each couple of vertices $(i,j)\in \Sig\times \Sig$ an edge $\ve_{ij}$ is an element of $E(\vI_{MC})$ iff
the probability of the transition $i\ra j$ is different from zero. The edge $\ve_{ij}$ is directed from $i$ to $j$.
\end{itemize}
\edefi

\brem
Note that in general transition probabilities are functions of $\bx$ and therefore in general time-dependent. This implies again the edge set $E$ of  $\vI_{MC}$ may change in time.
 \erem

\noindent In many situations it is more useful to describe a MC through  its infinitesimal generators rather than by the transition probabilities. The infinitesimal generator of a MC is determined by the \emph{rates} at which transitions occur. The rates themselves correspond to probabilities restricted to an infinitesimal time interval (see \cite{Stroock}). Let $\Sig = (\sig_1,...,\sig_M)$ be the states of the MC under consideration. At time $t$ the state of the MC is determined according to  the probability distribution

$$p(t)=(p_1(t),...,p_M(t)),$$ 

\noi where $p_i(t)$ is the probability that at time $t$ the system is in state $\sig_i$.  The respective normalisation condition is given by

$$\sum_{i=1}^Mp_i(t)=1.$$

\noi Given two states $i,j$, the transition $i\ra j$ occurring in a time interval $\delta t$ is determined by the probability 

$$P_{ij}=\left\{\begin{array}{ll}
\Kd_{ij}\de t\mbox{ for $i\neq j$},\\[4mm]
1-\de t\,\sum_{l\neq i}\Kd_{il}\mbox{ for $i= j$,}
\end{array}\right.$$

where $\Kd_{ij}$ is the $ij$-th element of the infinitesimal generator. Letting $\de t\ra 0$ the infinitesimal form of the dynamics can be written in matrix form:

\begin{equation}
\label{eq:MC-syst}
\dot{p}_i(t)=\sum_{j=1}^Mp_j(t)\,\Kd_{ji}~~i=1,...,M.
\end{equation}

{\remark
In general infinitesimal generators  depend on various parameters which might have their own dynamics and time dependence. We are interested in autonomous systems where the dynamics associated to the Markov chain transitions are all faster than the rest of the (in this paper mostly deterministic) dynamics which then may affect $\Kd$ on a longer time scale. Therefore we assume $\Kd$ not to be explicitly dependent on time, only implicitly via a possible dependence on $\bx$.} \\

\noi We are now able to state an alternative definition of $\vI_{MC}$ in terms of the infinitesimal generator $\Kd$:

\bdefi[The interaction graph $\vI_{MC}$ for Markov chain dynamics based on $\Kd$]
Given (\ref{eq:MC-syst}), the MC  \emph{interaction graph}  is a directed graph denoted by $\vI_{MC}=(\Sig,E)$, with $V(\vI_{MC}) = \{\sig_1, \ldots, \sig_M \}$ as before, and

\begin{itemize}
\item[(i)] an edge (arrow) $\ve_{ij}$  is an element of $E$ iff the couple of vertices $(i,j)$ (with $i\neq j$) is such that $\Kd_{ij}$ is not identically zero.
Moreover for every $i\in \Sig$, $\ve_{ii}$ is not in the set $E$.

\item[(ii)] the edge $\ve_{ij}$ is directed from $i$ to $j$ and is also denoted by $i\ra j$.
\end{itemize}
\edefi

\brem From the normalisation condition one can show (see \cite{Stroock}) that the  infinitesimal generator satisfies the following condition:

\begin{equation}
\label{cond:K}
\sum_{j=1}^M\Kd_{ij}=0\mbox{ for every $i=1,\ldots, M$.}
\end{equation}
\erem

\brem
Note that using $\Kd$ for $\vI_{MC}$ we can readily construct the 
\emph{adjacency} matrix $A$ by defining:
\[A_{ij}=\left\{\begin{array}{ll}
1 \mbox{ if $\Kd_{ij}\neq 0$}\\
0\mbox{ otherwise}
\end{array}\right.\]
\erem

\noindent Condition (\ref{cond:K}) implies that $\Kd$ has a non-trivial  left and right kernel. This is one of the main properties characterising  a MC. 

\brem
It holds that $\mu=(\mu_1,..,\mu_M)$ is a \emph{stationary measure} if

\begin{equation}
\sum_{j=1}^M\mu_j\Kd_{ji}=0\mbox{ for every $i=1,\ldots, M$},
\label{eq:stat-measure}
\end{equation}

\noi and

$$\sum_{j=1}^M\mu_j=1.$$
If $\mu_j>0$ for all $j$, then $\mu$  is said to be \emph{strictly positive}. 
\erem

For the spectrum of $\Kd$ we recall the following result:

\blem
The spectrum of  $\Kd$ is contained in the left half of the complex plane. 
\elem{\label{spectrum}}
This result can be proved by using Gershgorin's theorem (see \cite{MM}).

\subsubsection{Convex combinations of stationary measures}

Equation (\ref{eq:stat-measure}) defines the left kernel  of the matrix $\Kd$ (or the kernel of its transpose $\Kd^T$).  We denote the left kernel of $\Kd$ by $\ker(\Kd)$ and  assume that $\dim(\ker(\Kd))=L\geq 1$, namely $\ker(\Kd)$ is generated by $L$ stationary measures $\{\mu_l\}_{l=1}^L$. Under this condition we define

\bdefi[Convex combinations of stationary measures]  
Let $\{\mu^{(l)}\}_{l=1}^L$ be the stationary measures generating $\ker(\Kd)$. Let $\{\theta_l\}_{l=1}^L$ be $L$ positive real numbers with $\sum_{l=1}^L\theta_l=1$. We call the following sum

$$\mu=\sum_{l=1}^L\theta_l\mu^{(l)}$$ 

the convex combination of stationary measures associated to $\ker(\Kd)$.
\edefi

\noindent The convex combinations have the following properties:

\begin{enumerate}
\item Any convex combination is a probability measure.

\item Any convex combination is a stationary measure for the MC.
\end{enumerate}

\noindent Stationary measures are essential to describe the asymptotic behaviour of the MC. In fact one can show

\bth
Given any initial probability measure $p_0=(p_1(0),...,p_M(0))$,  there exist  $\{\theta_l\}_{l=1}^L$ with $\sum_{l=1}^L\theta_l=1$ such that
the solution $p(t)$ of (\ref{eq:MC-syst}) with $p(0)=p_0$ satisfies

$$\lim_{t\ra +\infty}p(t)=\mu$$

\noi with

$$\mu=\sum_{l=1}^L\theta_l\mu^{(l)}.$$ 
\noi The measure $\mu$ is called \emph{invariant measure}.
\eth

\bpf

This theorem rephrases some known results that can be found for example in Stroock \cite{Stroock}:

\begin{itemize}
 \item The matrix $\Kd$ generates a transition probability matrix by
\[p(t)=\mu_0\,P(t)\mbox{ where }P(t)=\exp(\Kd\,t),\]
where $\mu_0$ is the initial probability distribution.
\item Invariant measures $\mu$ are left-eigenvector of $P$ with eigenvalue $1$ and $\mu$'s 
form a convex set.
\item The space $\Sigma$ can partitioned into equivalence classes of states
\[C=\{i,j\in\Sigma:P_{ij}(t)>0,\,P_{ji}(t)>0\mbox{ for some $t\geq 0$}\}.\]
\item The limit $\lim_{t\ra +\infty}p(t)$ converges to an invariant measure 
 supported on the union of the equivalence classes $C$.
\end{itemize}

If the number of equivalence classes is $L$ a general invariant measure can be written as

$$\mu=\sum_{l=1}^L\theta_l\mu^{(l)}.$$ 

\epf

The preceding result shows that the asymptotic behaviour for $t\ra +\infty$ of a  Markov chain is classified by the stationary measures, namely by the solutions of equation (\ref{eq:stat-measure}). In the next section we shall describe the connections between the ergodic properties and the interaction graph in Markov chains.

\subsection{Interaction graph $\vI_{MC}$ and some ergodic properties} 
\label{subsec:mc}

In this paragraph we consider the ergodic properties of one single indecomposable Markov chain in terms of connectivity of its associated interaction graph. This indecomposable Markov chain will be associated to one single molecular machine being present. In case several such machines are present (as discussed later) the Cartesian product of state spaces gives rise to several disconnected components or subgraphs (which themselves of course should be internally connected in order to represent a meaningful single molecular machine). Such subgraphs can be discussed independently of each other, so without loss of generality we assume in this section there is only one connected component present. Let therefore $(\Sig,\Kd)$ be a Markov chain with (an unstructured) state space $\Sig$ and generator $\Kd$. Let us define

\bdefi
A directed path $i\Rightarrow j$ in $\vI_{MC}$ is a sequence $\{i_1,...i_n\}\subset \Sig$ such that
 
 $$i\ra i_1\ra i_2\ra...\ra i_n\ra j.$$ 
\edefi

\bdefi
A vertex $i$ is said to be connected to a vertex $j$ if $i\Rightarrow j$.
\edefi

In the theory of directed graphs (see for example \cite{Gibbons})  there are two possible definitions of connectivity. Let $\Sig$ be a non-empty set and consider the directed graph $(\Sig,E)$:

\begin{itemize}
\item[] $(\Sig,E)$ is a \emph{weakly connected graph} if for every $i,j\in \Sig$ at least one of the paths 
$i\Ra j$, $j\Ra i$ is not empty. 
\item[] $(\Sig,E)$ is a \emph{strongly connected graph} if for every $i,j\in \Sig$ both the paths 
$i\Ra j$, $j\Ra i$ are not empty. 
\end{itemize}

One can observe that a strongly connected $\vI_{MC}$ represents a $MC$ with a unique invariant measure $\mu$, with $\mu_i>0$ for all $i\in \Sig$. A weakly connected $\vI_{MC}$ corresponds to a MC with a unique $\mu$, but with possibly  some zero components. The reason for this outcome is that invariant measures tend to concentrate in certain vertices which are \emph{ends} of directed paths. This motivates the definition of \emph{end-vertices}:

\bdefi
A vertex $i_e\in \Sig$ is said to be an \emph{end-vertex} if

\begin{itemize}
\item there exists $i\in \Sig$ such that $i\Ra i_e$,
\item and there is no $j\in \Sig$ such that $i_e\Ra j$.
\end{itemize}
\edefi

\brem
In the context of Markov chain theory $i_e$ is an \emph{absorbing state} (see \cite{WC}). An \emph{absorbing} Markov chain contains \emph{at least} one \emph{absorbing} state (see \cite{WC}).
\erem

Now we propose to look at connected $\vI_{MC}$ which have at most one end-vertex. Such graphs will correspond to Markov chains with unique invariant measure that need not be necessarily strictly positive:
 
\bdefi
The graph $\vI_{MC}$ is called $\star$-connected if 

\begin{itemize}
\item $\vI_{MC}$ is connected,
\item $\vI_{MC}$ contains at most one end-vertex (absorbing state).
\end{itemize}
\edefi

\brem
Note that if a graph is \emph{strongly connected} then it is \emph{$\star$-connected}. If it is \emph{$\star$-connected},  then it is necessarily  \emph{weakly connected}.
\erem

\bprop
If $(\Sig,\Kd)$ is such that $\vI_{MC}$ is $\star$-connected, then $(\Sig,\Kd)$ has a unique invariant measure.
\eprop

\bpf
It suffices to show that a $\star$-connected $\Sig$ with $M$ states has a generator $\Kd$ with $\rank(\Kd)=M-1$. In the proof we shall construct interaction graphs by adding a vertex to a given graph. This process must be done without altering the property of being $\star$-connected, i.e. without increasing the number of end-vertices. We proceed by induction on $M$. For $M=2$, the graph is 

\[1\ra 2,\]

with  generator

\[\Kd_2=\left(\begin{array}{cc}
             k_{11} & k_{12}\\
		0 & 0
            \end{array}\right)
~~~\mbox{and  $k_{11}=-k_{12}$.}\]

Then one can readily show that

\[\rank{(\Kd_2)}=1.\]

The addition of a vertex leads to

\[1\ra 2 \ra 3\]

\[\Kd_3=\left(\begin{array}{ccc}
             k_{11} & k_{12} & 0\\
		0 & k_{22} & k_{23}\\
		0 & 0 & 0  
            \end{array}\right),
~~~\mbox{with $k_{11}=-k_{12}$, $k_{22}=-k_{23}$.}\] 

The rank of $\Kd_3$ is equal to $2$ since $\rank(\Kd_2)=1$, i.e. $k_{12}\neq 0$.
Consider $(\Sig_{M-1},\Kd_{M-1})$, with $M-1$ vertices and $\rank(\Kd_{M-1})=M-2$. The operation of adding a vertex leads to a new $\Sig_M$ with $M$ vertices and a generator

\[\Kd_M=\left(\begin{array}{cc}
                \Kd_{M-1} & k \\
		0 & 0
            \end{array}\right),  
~~~\mbox{with $k$ a column vector of dimension $M-1$.}\]

The vector $k$ gives the rates of the transitions from the vertices in $\Sig_{M-1}$ to the added $M^{th}$ vertex, hence $k$ is different from the zero vector. Up to re-numbering the vertices we can consider in $\Sig_M$ the vertex $M-1$ linked to the vertex $M$, hence $k_{M-1\,M}\neq 0$ and necessarily $k_{M-1\,M-1}\neq 0$; therefore $\det(\Kd_{M-1})\neq 0$ yields $\rank(\Kd_M)=M-1$.
\epf

Note that there are stronger purely algebraic conditions for the uniqueness of an invariant measure. The previous 'graphical' result can also be shown to be a simple  consequence of

\bprop
$\lambda$ is an eigenvalue of $\Kd$ if and only if it is an eigenvalue 
of a submatrix associated to a strongly connected component of the interaction graph associated to $\Kd$. 
\eprop

For the well-known proof of the proposition see for example \cite{WC}. For $\star$-connected graphs the unique end-vertex forms itself a strongly connected component that yields a zero eigenvalue which is therefore unique, and so leading to 
$\rank(\Kd_M)=M-1$.\\

The two interaction graphs  $\vI_{D}$ and $\vI_{MC}$ introduced so far can be interpreted as describing two different scales of the processes which now need to be related in a further modelling step. The deterministic interaction described by  $\vI_{D}$ is typically describing macroscopic properties, such as the overall 'systems' concentrations of different species of very abundant small molecules. Often in modelling these continuous variables forming the nodes of $\vI_{D}$ will be used for 'communication' between the states of the machinery described by the Markov chain. In many cases, such as in stirred chemical tank reactors there is no need to introduce any further discrete states for the system at all. The process description can be closed by introducing a set of all possible chemical reactions and -- by using mass action kinetics -- to derive the respective time-continuous dynamical system describing the chemical reactor macroscopically. The states of the Markov chain are connected to 'system switches' of one or several machines govering the processes. In our examples in the current paper such machines will be genes. They do not form concentrations of any kind that would need to be modelled in order to close the model description.

\section{The averaging principle, combined and averaged interaction graphs}

We now introduce the \emph{averaging principle} which has been derived  in \cite{LM1}. It can be interpreted as applying adiabatic theory to stochastic systems described by Fokker-Planck equations. Let us consider a complex system with hybrid dynamics whose state is fully determined by

\begin{enumerate}
\item $N$ real (continuous) variables $\bx=(x_1,...,x_N)\in\R^N$,

\item $M$ (discrete) states of a Markov chain.
\end{enumerate} 

\noindent At each time $t$ the state of the system is given by $(\bx,\sig)\in\R^N\times \Sig,$ where $\Sig=\{1,...,M\}$.  The dynamics of the MC is given by the specification of the rates, i.e. by specifying the infinitesimal generator $\Kd$. Therefore the probability of being in state $i$ at time $t$ is determined by the Cauchy problem

\begin{equation}
\label{MC-eq}
\left\{\begin{array}{ll}
\dot{p}_i(t)=\sum_{j=1}^Mp_j(t)\,\Kd_{ji}(\bx),\\
p_i(0)=p_{i,0},
\end{array}\right.
\end{equation}

\noi where in general the infinitesimal generator depends on $\bx$. In the following we make the basic assumption that the rates $\Kd_{ij}(\cdot)$ are continuously differentiable with respect to each $x_i$, $i=1,\ldots,N$.  If we fix a certain state $\sig$ of the MC, the time evolution of $\bx$ is determined by a set of vector fields $f^{(\sig)}(\bx)$ depending on the MC state. Therefore $\bx(t)=(x_1(t),...,x_N(t))$ is given by solving:

\begin{equation}
\label{det-eq}
\left\{\begin{array}{ll}
\displaystyle \frac{d^{(\sig)}{x}_i(t)}{dt}=f_i^{(\sig)}(x_1(t),...,x_N(t)), \mbox{ where $\sig=1,...,M$ and $i=1,...,N$,}\\[4mm]
x_i(0)=x_{i,0}.
\end{array}\right.
\end{equation}

\noi Here the superscript $(\cdot)^{(\sig)}$ denotes that the time derivative is determined by a vector field 
depending on the state $\sig$ of the MC. We are now in a state to associate an interaction graph to the combined hybrid model consisting of both continuous deterministic and discrete stochastic variables:

\bdefi
We define the \emph{Combined Interaction Graph}  $\vI_{C}=(V,E)$ by
$$ V(\vI_{C}) =  V(\vI_{D})  \cup  V(\vI_{MC})  = \{x_0, x_1, \ldots, x_N \} \cup \{\sigma_1,\ldots,\sigma_M \}, $$
$$ E(\vI_{C}) =  E(\vI_{D}) \cup E(\vI_{MC}) \cup E(\vI_{D} \to \vI_{MC}) \cup E(\vI_{MC} \to \vI_{D}), $$
with $(x_i, \sigma_j) \in E(\vI_{D} \to \vI_{MC})$  iff $ \frac{d}{d x_i} \Kd_{jj}(\bx) \neq 0$, with $i=0,\ldots,N$, $j=1,\ldots,M$, and $( \sigma_i, x_j) \in E(\vI_{MC} \to \vI_{D})$ iff $ \frac{d}{d x_j} f_j^{(\sig_i)}(\bx) \neq 0$, with $i=0,\ldots,M$ and $j=0,\ldots,N$.
\edefi

This means the combined interaction graph  vertex set has as expected the union of the vertex sets of the deterministic interaction graph $\vI_{D}$ and the Markov chain interaction graph $\vI_{MC}$. The edge set of the new hybrid graph $\vI_{C}$ also contains the union of the edge sets of  $\vI_{D}$ and $\vI_{MC}$, but  $E(\vI_{C})$ is possibly larger. The set $E(\vI_{D} \to \vI_{MC})$ describes how the continuous variables influence the switching rates of the Markov chain, whereas $E(\vI_{MC} \to \vI_{D})$ describes how states of the Markov chain influence the continuous variables. In our examples of $\vI_{C}$ (Fig. \ref{fig:OR_gate}, Fig. \ref{fig:AND_gate}, Fig. \ref{fig:BL_cc_graph}) edges contained in $E(\vI_{MC} \to \vI_{D})$ are drawn by simple broken line arrows, whereas edges in $E(\vI_{D} \to \vI_{MC})$ are drawn by bold solid arrows. We also use the convention that if  a node in $V(\vI_{D})$ influences all the nodes of the MC, then the bold arrow is drawn only once to the box surrounding the Markov chain graph.  With these concepts we can  now conveniently introduce the \emph{adiabatic} or \emph{averaging} principle:

\subsubsection*{Averaging Principle and Averaged Interaction Graph}
Upon the assumption that the evolution of the MC is faster than the deterministic dynamics, the 
macroscopic evolution of the combined system is described by the average dynamics, which is  given by

\begin{equation}
\label{average-eq}
\left\{\begin{array}{ll}
\displaystyle \frac{d x_i(t)}{dt}=\sum_{\sig=1}^M\mu_\sig(\bx(t))\,f_i^{(\sig)}(\bx(t)),\\[4mm]
x_i(0)=x_{i,0}.
\end{array}\right.
\end{equation}

Here $\mu$ is  an invariant measure of $\Kd(\bx)$, and  $i=1,...,N$. The invariant measure $\mu$ can be a convex combination of invariant measures compatible with the initial values $\{p_i(0)\}_{i=1}^M$, namely 
the support of $\mu$ in $\Sig$ is equal to the support of $p(0)$.  Note that the average vector field can also be written in matrix compact form

\begin{equation}
\label{matrix-average-eq}
\dot\bx(t)=\mu(\bx(t))\,F(\bx(t)),
\end{equation}

\noi where $\mu(\bx)=(\mu_1(\bx),...,\mu_M(\bx))$ and $F(\bx)=\{f_i^{(\sig)}(\bx)\}_{i=1,...N}^{\sig=1,...,M}$.

\bdefi
We define the \emph{Averaged Interaction Graph}  $\bar{\vI}=(V,E)$ by $V(\bar{\vI}) =  V(\vI_{D}) $ and $E(\vI) =  E(\vI_{D}) \cup E_{\downarrow}$, with $(x_i, x_j) \in E_{\downarrow}$, $i,j = 1,\ldots, N$  iff 
$$ \frac{\pa}{\pa x_i}\sum_{\sig=1}^M\mu_\sig(\bx(t))\,f_j^{(\sig)}(\bx(t)) \neq 0. $$
\edefi

This essentially means that all (microscopic) interactions of the MC visualised by nodes and edges  are lumped into edges only by applying the averaging principle, the edges of $E_{\downarrow}$. This is precisely illustrated in Fig. \ref{fig:system}. The edges in $E_{\downarrow}$ are drawn by convention as red(grey) broken line arrows, see the examples Fig. \ref{fig:OR_gate_averaged}, Fig. \ref{fig:AND_gate_averaged} and Fig. \ref{fig:BL_cc_graph-average}.

\brem
This is an important remark about feedback loops and qualitative behaviour of the average dynamics. Because $\bar{\vI}$ is again the interaction graph of a deterministic vector field, all the known results about the connection between directed loops or cycles contained in $\bar{\vI}$ and possible restrictions on the attractor sets apply. For example the classical result by Thomas-Soul\'e holds, see \cite{review}. To check the respective conditions it is necessary to look at sign weighted interaction graphs, i.e. the existence conditions for arrows need to be refined in the sense that the sign of the respective derivatives needs to be evaluated. This is simple in all cases of interaction graphs introduced in this paper. Excitation or stimulation symbols in application diagrams correspond to arrows with a positive weight, whereas arrows with negative weights correspond to inhibition symbols (compare with Fig. \ref{fig:BL_cc}). It is presently unclear in which cases the averaging produces monotone dynamical systems \cite{smith} which generically do not exhibit oscillatory behaviour (compare with section \ref{subsub:oscillations}), or when reversely positive feedbacks (positive directed cycles in $\bar{\vI}$) are created that are a condition for oscillatory behaviour by the  Thomas-Soul\'e theorem. Similar remarks could be made for all function classes, for example about the quasi-monotone functions, or the ocurence of vector fields allowing chaotic dynamics. In general the sign structure of the Jacobian or equivalently the weights of $\bar{\vI}$ are time- and therefore state dependent, with the monotone and quasi-monotone vector fields being exceptions.
 \erem

\subsection{Markov chain structure and the average dynamics}
\label{structure}

The MC structure is determined by $\Sig$ and by its infinitesimal generator $\Kd$. The state space $\Sig$ consists of all possible discrete states corresponding to those molecular (or other) states that cannot be described by continuous variables (i.e. concentrations). In the systems of interest there are as discussed usually in addition many molecular (or more generally 'discrete state') machines that must be modeled by a discrete space structure independently. Let  us now assume that there are more than one, say $M$ many such entities. Then each of them will have  its own state space $\Sig_i$ with $i=1,\ldots,M$. The total space is therefore the Cartesian product

\begin{equation}
\label{tot-Sig}
\Sig=\times_{i=1}\Sig_i.
\end{equation}

Each $\sig\in\Sig$ has the form $\sig=\sig_{i_1...i_M}\doteq (O_{1i_1},...,O_{Mi_M})$, with $O_{ji_j}\in\Sig_j$.
As explained in \cite{LM2}, the infinitesimal generator $\Kd$ is constructed by assuming that a transition  

$$\sig\ra\sig'$$

involves only one factor space $\Sig_j$, namely there exits only one index $i$ such that

$$\sig=(O_{1i_1},...,O_{ik_i},...,O_{Mi_M})\ra\sig'=(O_{1i_1},...,O_{ik'_i},...,O_{Mi_M}).$$

This can be called a principle of local effect and is very reasonable for molecular systems. Let us consider the simplest case which is $\Sig=\Sig_1\times\Sig_2$, where $\Sig_l=\{O_{l1},O_{l2}\}$,  $l=1,2$. Set $\sig_{i_1i_2}=(i_1i_2)$, where in order to simplify notation we identify the index with the state. The average dynamics then has the following form:
 
 $$\dot{\bx}(t)=\sum_{(i_1i_2)\in\Sig_1\times\Sig_2}\mu_{(i_1i_2)}(\bx(t))\,f^{(i_1i_2)}(\bx(t))=
 \sum_{i_1\in\Sig_1,i_2\in\Sig_2}\mu_{(i_1i_2)}(\bx(t))\,f^{(i_1i_2)}(\bx(t)),$$
 
 where
 
 \begin{equation}
 \sum_{(i_1i_2)\in\Sig_1\times\Sig_2}\mu_{(i_1i_2)}(\bx)=1
 ~~\forall \bx\in\R^N.
 \label{normalisation}
 \end{equation}
 
Now suppose that the vector field is a linear combination of the form

$$f^{(i_1i_2)}(\bx)=\alpha_1\,g^{(i_1)}(\bx)
+\alpha_2\,g^{(i_2)}(\bx)~~~\forall\bx\in\R^N,~~~\forall (i_1,i_2)\in\Sig_1\times\Sig_2,$$

with $\alpha_1,\alpha_2\in\R$. Then the average vector field turns out to be

\begin{equation}
\dot{\bx}(t)=\alpha_1\cX_1(\bx(t))+\alpha_2\cX_2(\bx(t)),
\end{equation}

where the vector fields $\cX_1$ and $\cX_2$ are given by

\begin{equation}
\begin{array}{ll}
\displaystyle\cX_1(\bx)=\sum_{i_1\in\Sig_1}\left(\sum_{i_2\in\Sig_2}
\mu_{(i_1i_2)}(\bx)\right)\,g^{(i_1)}(\bx),  \\[5mm]
\displaystyle\cX_2(\bx)=\sum_{i_2\in\Sig_2}\left(\sum_{i_1\in\Sig_1}
\mu_{(i_1i_2)}(\bx)\right)\,g^{(i_2)}(\bx).\\
\end{array}
\label{vcX}
\end{equation}

Note that the normalisation (\ref{normalisation}) implies that the two expressions

\begin{equation}
\begin{array}{ll}
\displaystyle\mu_{1,(i_1)}(\bx)\doteq\sum_{i_2\in\Sig_2}\mu_{(i_1i_2)}(\bx),\\[5mm]
\displaystyle\mu_{2,(i_2)}(\bx)\doteq\sum_{i_1\in\Sig_1}\mu_{(i_1i_2)}(\bx)
\end{array}
\label{prob}
\end{equation}

form two probability measures respectively over $\Sig_1$ and $\Sig_2$. As discussed in \cite{LM2} the transpose of the generator $\Kd$ has components that can be written as

\begin{equation}
k_{(ij);(i'j')}=\left\{\begin{array}{lll}
k_{1,(ii')}\mbox{ if $j=j'$},\\[3mm]
k_{2,(jj')}\mbox{ if $i=i'$},\\[3mm]
0\mbox{ if $i\neq i'$ and $j\neq j',$}
\end{array}\right.
\label{rates-for-product}
\end{equation}

where $k_{1,(.)}$ and $k_{2,(.)}$ are the entries of the matrices  $\Kd_1,\Kd_2$ associated to the 
MC on $\Sig_1$ and $\Sig_2$ respectively. 

\bprop
If the infinitesimal generator $\Kd$ satisfies relation (\ref{rates-for-product})  then  the measures $\mu_{1}(\bx)$ and $\mu_2(\bx)$ are invariant measures of the MC's $(\Kd_1,\Sig_1)$  and $(\Kd_2,\Sig_2)$, respectively.
\eprop

\bpf
It is sufficient to prove the result for one MC. Let us compute

$$\sum_{i_1'\in\Sig_1}k_{(1,i_1i'_1)}\mu_{1,(i_1)}.$$

This is equal to

$$\sum_{i_1'\in\Sig_1}\sum_{i_2\in\Sig_2}k_{(1,i_1i'_1)}\mu_{(i'_1,i_2)}.$$

The last expression can be rewritten as

\begin{equation}
\sum_{i_1'\in\Sig_1}\sum_{i_2\in\Sig_2}
\sum_{i'_2\in\Sig_2}k_{(1,i_1i'_1)}\de_{i_2i_2'}\mu_{(i'_1,i'_2)}.
\label{form0}
\end{equation}

Now using the definition (\ref{rates-for-product}),  expression (\ref{form0}) becomes

\begin{equation}
\sum_{i_1'\in\Sig_1}\sum_{i_2\in\Sig_2}
\sum_{i'_2\in\Sig_2}k_{(i_1i'_1);(i_2i_2')}\mu_{(i'_1,i'_2)}=0.
\label{form1}
\end{equation}

The latter equation is satisfied because $\mu$ is an invariant measure on $(\Kd,\Sig_1\times\Sig_2)$, and therefore

$$ \sum_{i_1'\in\Sig_1}
\sum_{i'_2\in\Sig_2}k_{(i_1i'_1);(i_2i_2')}\mu_{(i'_1,i'_2)}=0.$$
\epf

The result of this paragraph can be easily generalised to systems where the MC is constructed by the product  of $M$ MC's. We leave this extension to the reader.

\section{Examples}
This section is devoted to different examples of increasing difficulty which will explain and demonstrate the approach and introduce examples of interaction graphs. The first two examples contains the analysis of a so-called feed-forward-loop (FFL) whose modelling has been introduced in \cite{MA}. In this context we look at both an AND and OR gate. We took these two examples to show that the average dynamics is a relatively natural way to attach testable realistic dynamics to a well known genetic wiring diagram, or mathematically speaking, to a directed graph. Our final example will present a new point of view on  modelling a genetic oscillator. In our models each single gene is described through a set of finite discrete states with the dynamics given by a Markov chain. The switching of genes is triggered by  molecules described as a continuum, typically interpreted as transcription factors that are sufficiently abundant to be described by a continuum variable. In this setting we show that the average dynamics for the clock allows to demonstrate circadian rhythms using a two-gene system  -- without the need to introduce the assumption that genes are modelled as 'gene concentrations'. This is unfortunately an underlying assumption for most existing models found in the literature. Note that in the modeling step we usually interpret transitions as reaction schemes, involving classical mass-action reaction for continuous variables, stochastic transitions for MCs, but also combined reactions incorporating both mass-action kinetics and stochastic transitions. See \cite{review},  \cite{LM1} for a more detailed introduction to this extended reaction scheme formalism. 

\subsection{Feed-forward loops}

\begin{figure}[htbp] 
   \centering
   \includegraphics[scale=0.4]{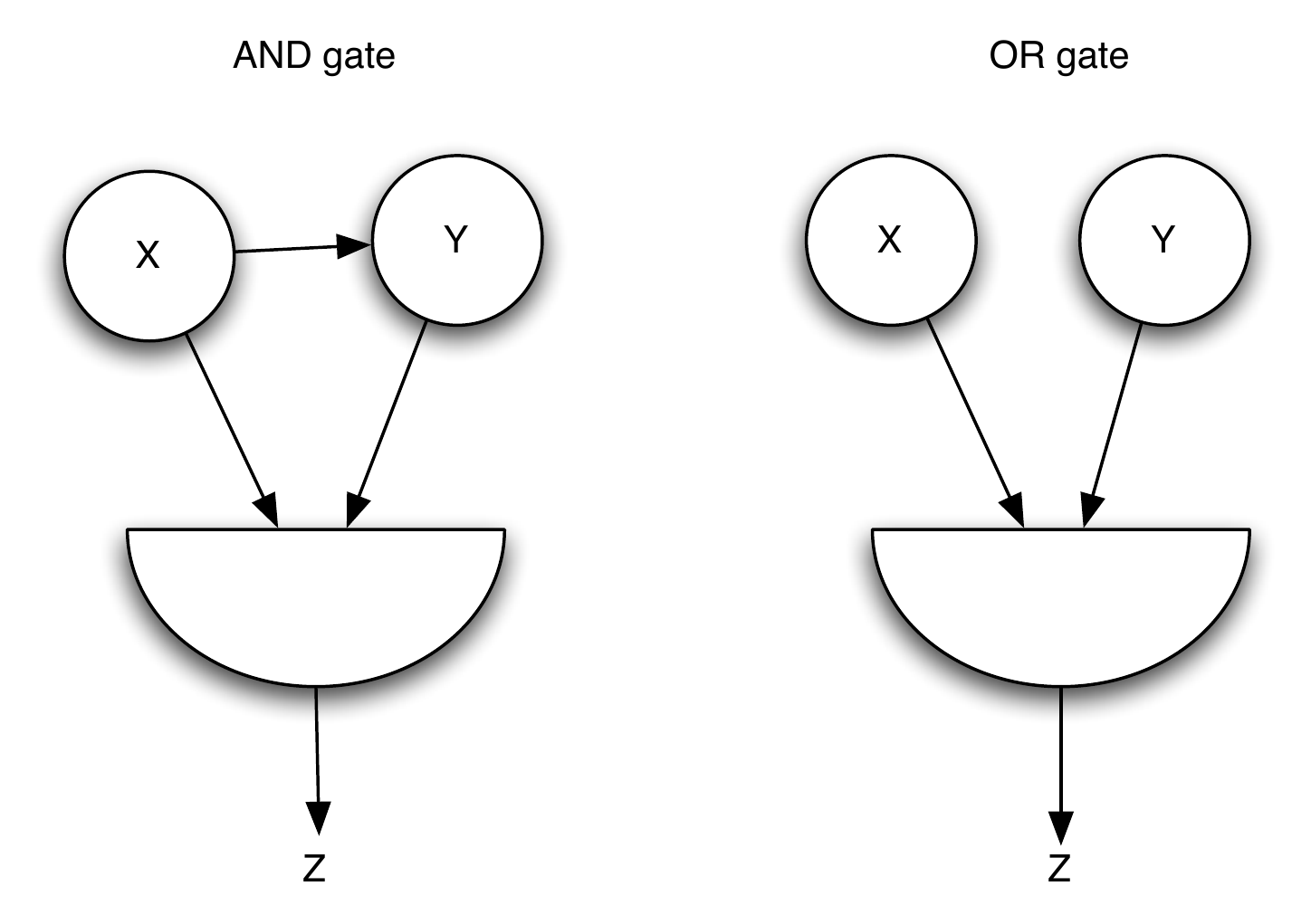} 
   \caption{AND gate and OR gate according to U. Alon et al.}
   \label{fig:Uri_Alon}
\end{figure}

In \cite{MA}, Mangan and Alon analysed  so called feed-forward loops (FFL) describing genetic circuits. We interpret them here as a single wiring diagram only, assumed to be independent from any other process occurring in a cell. This means we are not interpreting them as a small subgraphs of a larger graph, i.e. as so-called motifs. Such loops if isolated in this sense are generically described through diagrams like Figure \ref{fig:Uri_Alon}. In principle the choice of the associated dynamics is open, see the discussion in \cite{review}. One can first select the interaction graph, then the dynamics, or vice versa. There is in other words no one-to-one relationship between the graphical model and the equation determining the time evolution. A better interpretation of the interaction graph is that the feed-forward loop is determining a possible class of equations. The question is then whether this class has distinctive common features which can be determined - or not. The generic class of equations associated to the FFL in the literature is  chosen as a system of ordinary differential equations which associated vector field is compatible with the graphical representation of the FFL. In \cite{MA} the diagram in Fig.\ref{fig:Uri_Alon} is associated to a dynamics given by

\begin{equation}
\left\{\begin{array}{ll}
\dot{y}=b_y+\beta_y\,f(x^*,k_{xy})-\alpha_y\, y,\\[3mm]
\dot{z}=b_z+\beta_z\,G(x^*,k_{xz},y^*,k_{yz})-\alpha_z\, z,\\[3mm]
\end{array}\right.
\label{eq_alon}
\end{equation}

\noi where

$$x^*=\left\{\begin{array}{ll}
x\mbox{ if $Sx=1$,}\\
0\mbox{ if $Sx=0$.}
\end{array}\right.
$$

The $y^*$, the transcription factor $Sy$ and $b_y,b_z,\beta_y,\beta_z, \alpha_y$ and $ \alpha_z$ are all positive constants.  The key elements of (\ref{eq_alon}) are the non-linear functions $f$ and $G$. For an activator we assume

$$f(u,k)=\frac{(u/k)^H}{(1+(u/k)^H},$$

\noi whereas for a repressor the non-linearity becomes

$$f(u,k)=\frac{1}{(1+(u/k)^H}.$$

\noi The gate function for an AND-gate is given by

\begin{equation}
G=f(x^*,k_{xz})f(x^*,k_{yz}).
\label{AND-gate}
\end{equation}

\noi  If there are two transcription factors competing for binding to the promoter region,  a possible choice for  an activator is

$$f_c(u;k_u,k_v,v)=\frac{(u/k_u)^H}{1+(u/k)^H+(v/k)^H},$$

\noi and for a repressor

$$f_c(u;k_u,k_v,v)=\frac{1}{1+(u/k)^H+(v/k)^H}.$$

\noi For the OR-gate in \cite{MA} the $G$ \emph{transfer} function is

\begin{equation}
G=f_c(x^*;k_{xz},k_{yz},y^*)+f_c(x^*;k_{yz},k_{xz},y^*).
\label{OR-gate}
\end{equation}

\noi Next we demonstrate that the \emph{averaging principle} can be used to derive effective dynamics with qualitative properties similar to (\ref{AND-gate}) and (\ref{OR-gate}) giving a new interpretation and justification to the choice of non-linearities selected to describe the dynamics. The derivation is based on 
on a sequence of assumptions necessary to interpret  graphs like Fig. \ref{fig:Uri_Alon}. The interpretation is essentially related to the description of the processes at a semi-microscopic  level where  gene activation is modeled. Because assumptions are made on the microscopic scale, they are open to experimental (molecular) validation.

\subsubsection*{OR-gate}
We assume that $X$ and $Y$ are competing for a single binding site $P$. This scenario can be modelled by the following set of reactions:

\begin{equation}
\begin{array}{llll}
X+P\rlha[h_1]{k_1}PX\ra^{\beta_x}PX+ Z\\
Y+P\rlha[h_2]{k_2}PY\ra^{\beta_x} PY+Z\\
Z\ra^\de\emptyset\\
\emptyset\ra^{b_z} Z
\end{array}
\end{equation}

\begin{figure}[htbp] 
   \centering
   \includegraphics[scale=0.4]{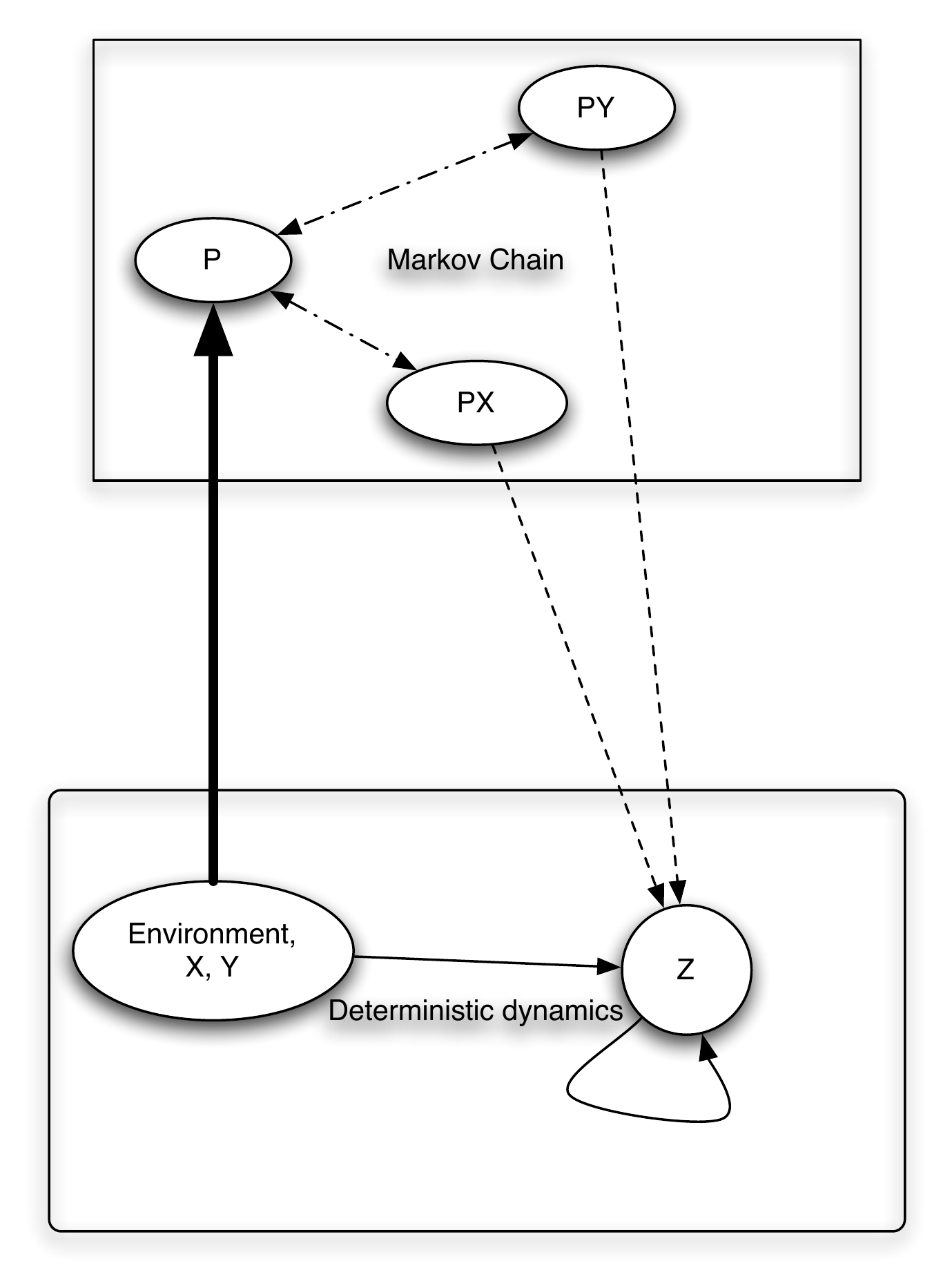} 
   \caption{The OR gate combined interaction graph $\vI_{C}$. Note that the environment node $x_0 = (x,y)$ has two components.}
   \label{fig:OR_gate}
\end{figure}

\noi The OR-gate reactions can be described using the graph in Fig. \ref{fig:OR_gate}. Note that in this case we need to include an \emph{environment} vertex. In the graphical representation the \emph{environment} vertex includes also the species $X$ and $Y$ which affect the MC but  do not have an explicit dynamics. The MC has the state space

$$\Sig=\{P,PX,PY\},$$

\noi and the infinitesimal generator is given by

$$\Kd=\left(\begin{array}{ccc}
-k_1x-k_2y & k_1x & k_2y\\
h_1 & -h_1 &0\\
h_2 & 0 & -h_2
\end{array}\right).$$

\noi The deterministic dynamics is chosen as simple as possible:

$$\begin{array}{lll}
\displaystyle \dot{z}(t)=b_z-\de\,z\mbox{ if the MC is in state $P$},\\
\displaystyle \dot{z}(t)=b_z-\de\,z+\beta_x\mbox{ if the MC is in state $PX$},\\
\displaystyle \dot{z}(t)=b_z-\de\,z+\beta_y\mbox{ if the MC is in state $PY$.}
\end{array}$$

\noi The invariant measure of the MC is

$$\mu(x,y)=\left({\frac {h_{{1}}h_{{2}}}{h_{{1}}h_{{2}}+k_{{1}}h_{{2}}\,x+k_{{2}}h_{{1}}\,y
}},{\frac {k_{{1}}h_{{2}}\,x}{h_{{1}}h_{{2}}+k_{{1}}h_{{2}}\,x+k_{{2}}h_
{{1}}\,y}},{\frac {k_{{2}}h_{{1}}\,y}{h_{{1}}h_{{2}}+k_{{1}}h_{{2}}\,+k_{{2}
}h_{{1}}\,y}}\right],$$

\noi and the average dynamics turns out to be

$$\dot{z}=b_z-\de\,z+\beta_x\,{\frac {k_{{1}}xh_{{2}}}{h_{{1}}h_{{2}}+k_{{1}}xh_{{2}}+k_{{2}}yh_
{{1}}}}+\beta_y\,{\frac {k_{{2}}yh_{{1}}}{h_{{1}}h_{{2}}+k_{{1}}xh_{{2}}+k_{{2}
}yh_{{1}}}}.$$

\noi The  \emph{transfer} function of the OR-gate then becomes
  
$$G=\beta_x\,{\frac {k_{{1}}xh_{{2}}}{h_{{1}}h_{{2}}+k_{{1}}xh_{{2}}+k_{{2}}yh_
{{1}}}}+\beta_y\,{\frac {k_{{2}}yh_{{1}}}{h_{{1}}h_{{2}}+k_{{1}}xh_{{2}}+k_{{2}
}yh_{{1}}}}.$$

\begin{figure}[htbp] 
   \centering
   \includegraphics[width=3in]{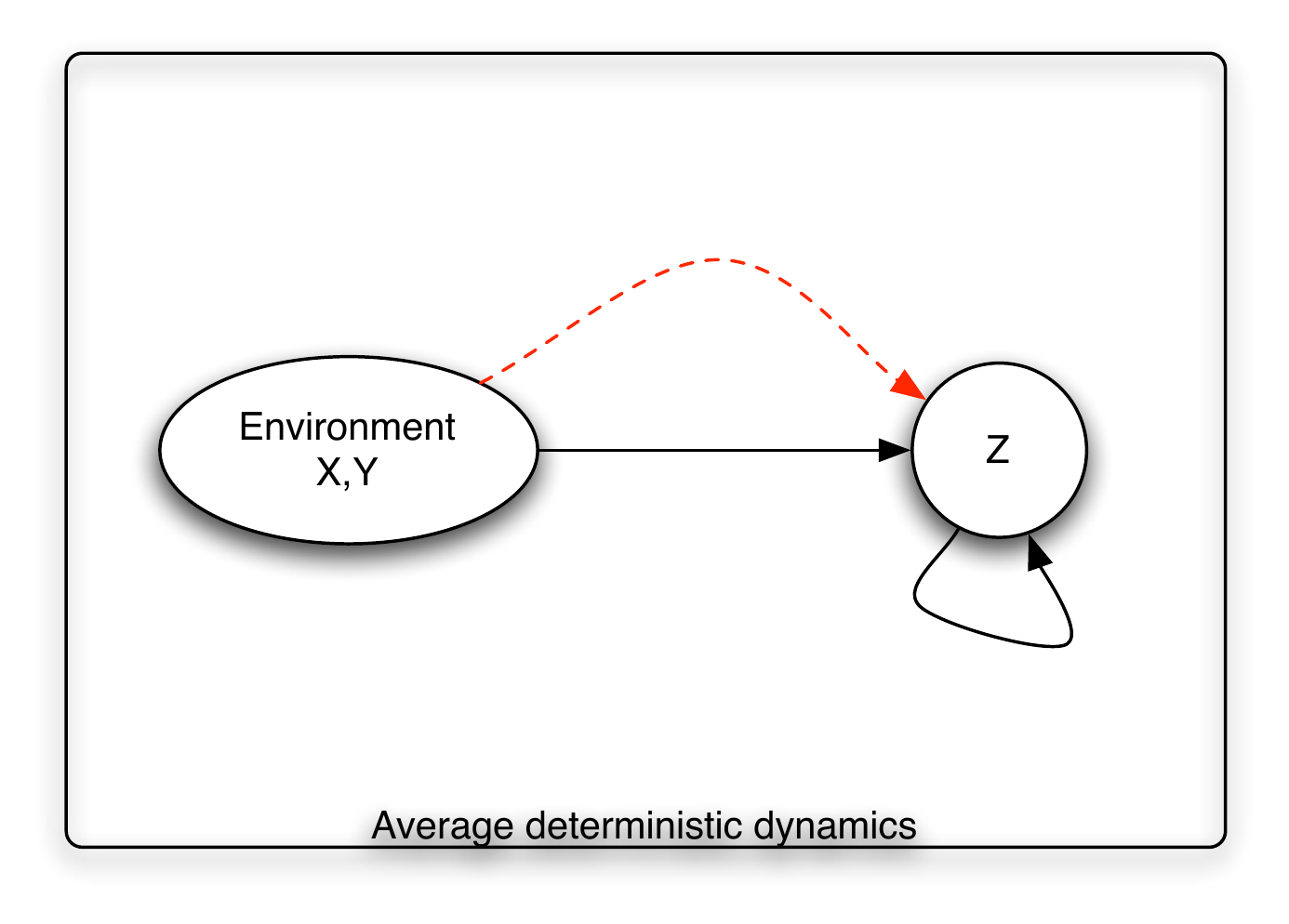} 
   \caption{ In the interaction graph $\bar{\vI}$   associated to the average dynamics the whole MC is substituted by  a second (henceforth \emph{new})  edge directed the environment state to $Z$.}
   \label{fig:OR_gate_averaged}
\end{figure}

\noi The average dynamics has an associated interaction graph $\bar{\vI}$ as visualised in
Fig. \ref{fig:OR_gate_averaged}.

\subsubsection*{AND-gate}
In this case we assume that  both $X$ and $Y$ can temporarily bind to  site $P$. This can be modelled with the following set of reactions:

\begin{equation}
\begin{array}{lll}
\mbox{$X$ regulates $Y$:}\\
X+P\rlha[h_1]{k_1}PX\\
PX\ra^{b_y} PX + Y
\end{array}
\qquad
\begin{array}{lllll}
\mbox{$X$ and $Y$ regulate $Z$:}\\
Y+P\rlha[h_2]{k_2}PY\\
X+PY\rlha[h_1]{k_1}PXY\\
Y+PX\rlha[h_2]{k_2}PXY\\
PXY\ra^\beta PXY+Z
\end{array}
\end{equation}
\begin{equation}
\begin{array}{lll}
\mbox{Degradation of $Y$ and $Z$:}\\
Z\ra^{\de_z}\emptyset\\
Y\ra^{\de_y}\emptyset
\end{array}
\qquad
\begin{array}{lll}
\mbox{Basal expression of $Y$ and $Z$:}\\
\emptyset\ra^{b_z} Z\\
\emptyset\ra^{b_y} Y
\end{array}
\end{equation}

\begin{figure}[htbp] 
   \centering
   \includegraphics[scale=0.4]{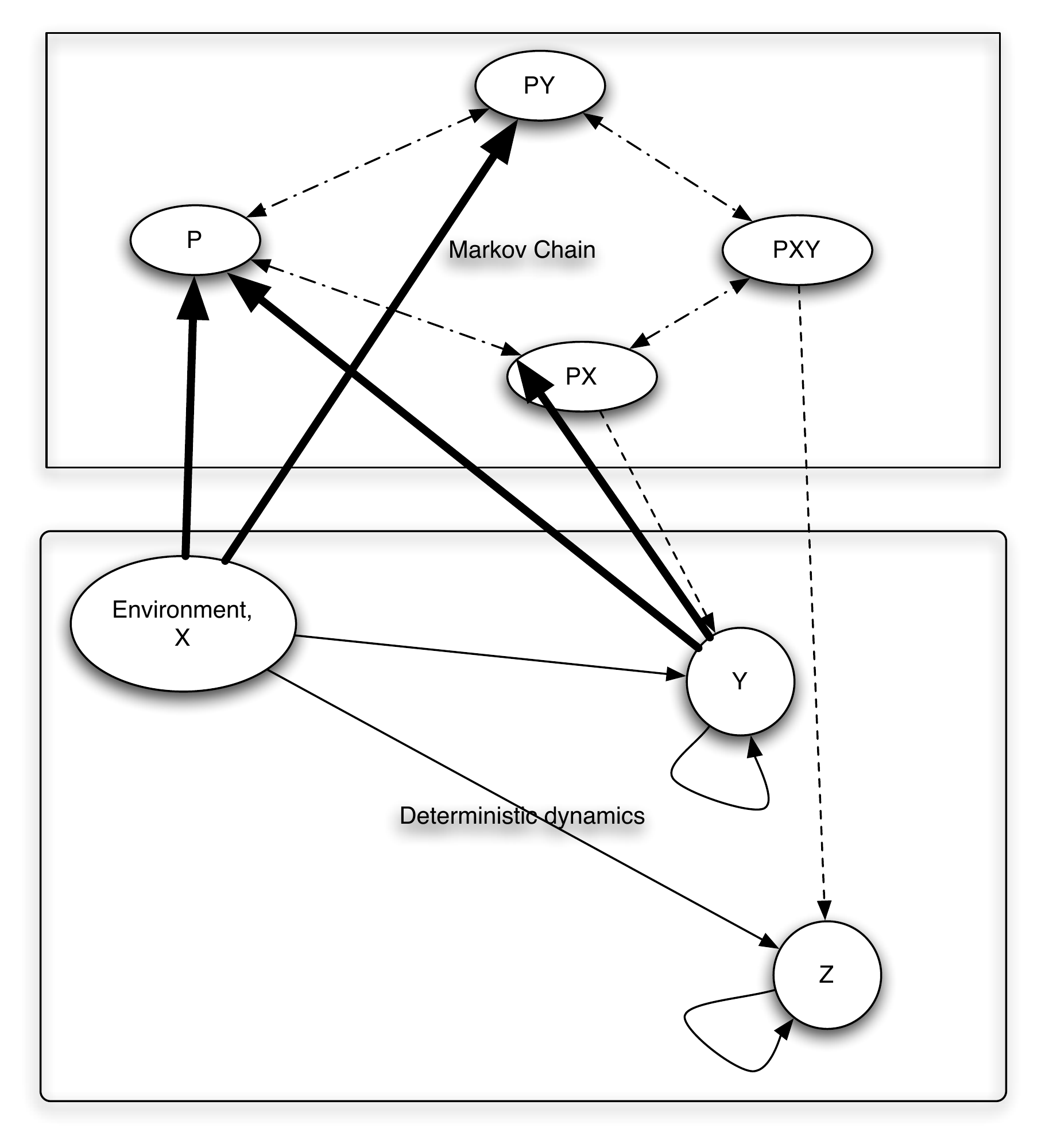} 
 \caption{The AND gate combined interaction graph $\vI_{C}$.}
   \label{fig:AND_gate}
\end{figure}

\noi These reactions can be again described through the interaction graph  given in  
Figure \ref{fig:AND_gate}. Also in this case an \emph{environment} vertex  has to be introduced. The Markov chain has state space

$$\Sig=\{P,PX,PY,PXY\}.$$

\noi The corresponding infinitesimal generator is given by

$$\Kd=\left(\begin{array}{cccc}
-k_1x-k_2y & k_1x & k_2y & 0\\
h_1 & -h_1-k_2y &0 & k_2y\\
h_2 & 0 & -h_2-k_1x & k_1x\\
0 & h_2 & h_1 & -h_1-h_2
\end{array}\right).$$

\noi  The deterministic dynamics is again chosen as simple as possible:

$$\begin{array}{llll}
\displaystyle \dot{y}(t)=b_y-\de_y\,y\mbox{ if the MC is the state $P$, $PY$ and $PXY$},\\[2mm]
\displaystyle \dot{y}(t)=b_y-\de_y\,y+\beta_y\mbox{ if the MC is the state $PX$},\\[2mm]
\displaystyle \dot{z}(t)=b_z-\de_z\,z\mbox{ if the MC is the state $P$, $PX$ and $PY$},\\[2mm]
\displaystyle \dot{z}(t)=b_z-\de_z\,z+\beta_z\mbox{ if the MC is the state $PXY$.}
\end{array}$$

\noi The invariant measure of the MC is

$$\begin{array}{ll}
\mu(x,y)=\displaystyle\left({\frac {h_{{2}}h_{{1}}}{h_{{2}}h_{{1}}+k_{{1}}xh_{{2}}+k_{{2}}yh_{{1}
}+k_{{1}}xk_{{2}}y}},{\frac {k_{{1}}xh_{{2}}}{h_{{2}}h_{{1}}+k_{{1}}xh
_{{2}}+k_{{2}}yh_{{1}}+k_{{1}}xk_{{2}}y}},\right.\\[4mm]
\displaystyle\left. {\frac {k_{{2}}yh_{{1}}}{h_{
{2}}h_{{1}}+k_{{1}}xh_{{2}}+k_{{2}}yh_{{1}}+k_{{1}}xk_{{2}}y}},{\frac 
{k_{{1}}xk_{{2}}y}{h_{{2}}h_{{1}}+k_{{1}}xh_{{2}}+k_{{2}}yh_{{1}}+k_{{
1}}xk_{{2}}y}}\right),
\end{array}$$

\noi and the average dynamics turns out to be

$$\left\{\begin{array}{ll}
\displaystyle \dot{y}=b_y-\de_y\,y+\beta_y\,{\frac {k_{{1}}xh_{{2}}}{h_{{2}}h_{{1}}+k_{{1}}xh
_{{2}}+k_{{2}}yh_{{1}}+k_{{1}}xk_{{2}}y}},\\[4mm]
\displaystyle \dot{z}=b_z-\de_z\,z+\beta_z\,{\frac 
{k_{{1}}xk_{{2}}y}{h_{{2}}h_{{1}}+k_{{1}}xh_{{2}}+k_{{2}}yh_{{1}}+k_{{
1}}xk_{{2}}y}}.
\end{array}\right.$$

\noi This implies we obtain an AND-gate with \emph{transfer} function

$$G=\beta\,{\frac 
{k_{{1}}xk_{{2}}y}{h_{{2}}h_{{1}}+k_{{1}}xh_{{2}}+k_{{2}}yh_{{1}}+k_{{
1}}xk_{{2}}y}}.$$

\begin{figure}[htbp] 
   \centering
   \includegraphics[scale=0.3]{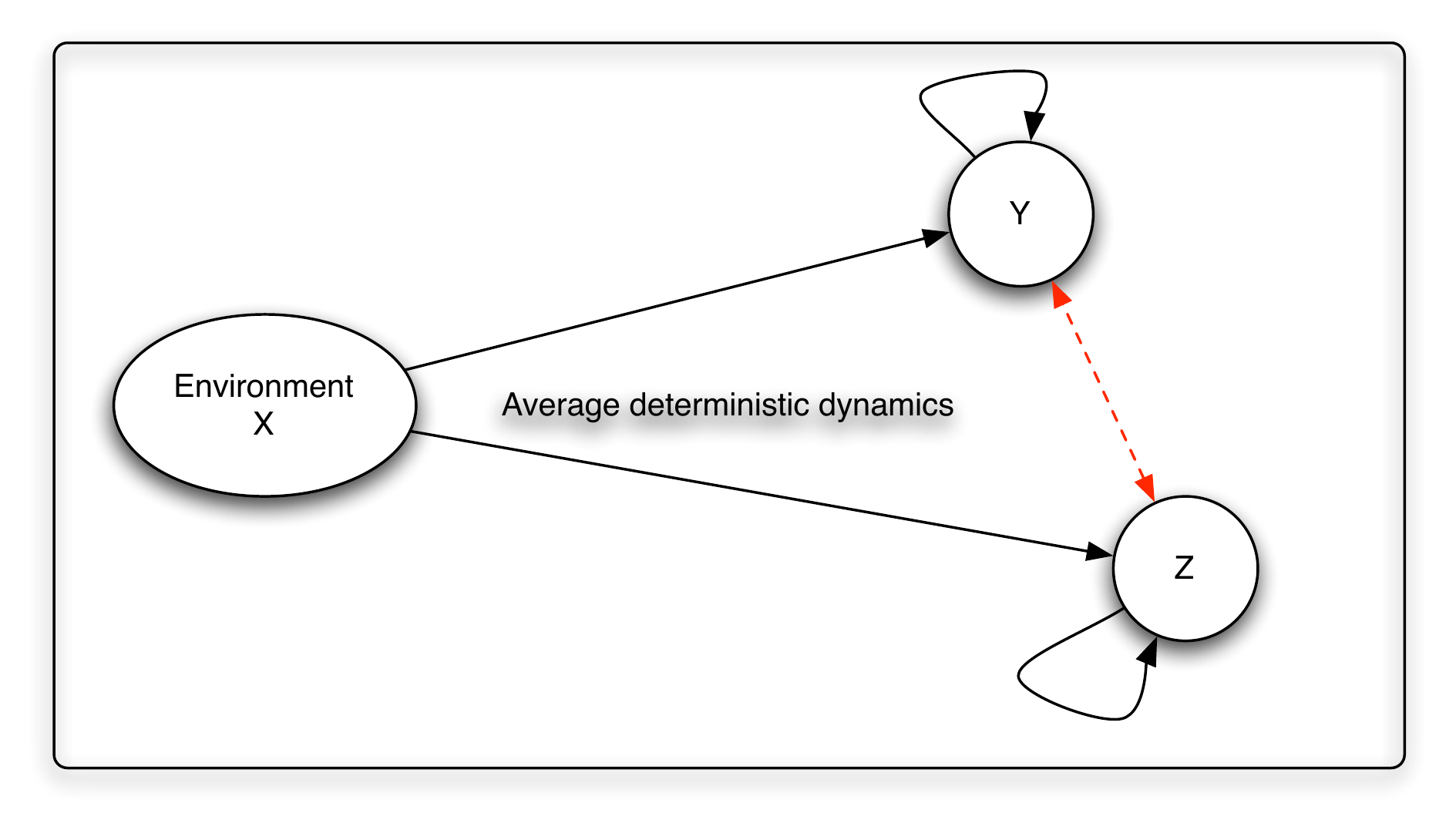} 
   \caption{The averaged interaction graph $\bar{\vI}$  associated to the average dynamics of the AND gate.  Two new edges are formed between $Y$ and $Z$. These links correspond to new effective interactions as described by the edge set $E_{\downarrow}$.}
   \label{fig:AND_gate_averaged}
\end{figure}

\noi The averaged dynamics has an associated interaction graph $\bar{\vI}$  as given in Fig. \ref{fig:AND_gate_averaged}. 

\subsection{An application to genetic oscillators}
\begin{figure}[htbp] 
   \centering
   \includegraphics[scale=0.4, width=7cm]{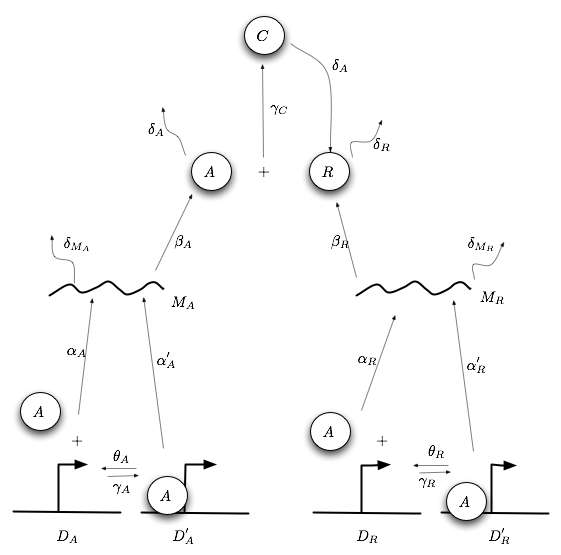} 
   \caption{Genetic Oscillator Model by N. Barkai et al.}
   \label{fig:BL_cc}
\end{figure}

Cyclic behaviours are important elements for the understanding of regulation inside any biological process. It is no surprise that therefore oscillations  are at the centre of many different approaches to their modelling and analysis (see \cite{atkinson,elowitz,gardner,hasty,Amos,VKBL,precision,2comp}).  We draw particular attention to \cite{VKBL} where a genetic oscillator using a system of two types of interacting genes has been designed. The structure of the model is presented in Fig. \ref{fig:BL_cc}.  The two genes $gene_A$ and $gene_R$ express two proteins $A$ (activator) and $R$ (repressor). The activator $A$ enhances $gene_A$ and $gene_R$.  The activator $A$ inhibits the repressor $R$ by forming the complex $C$. In \cite{VKBL} the model is constructed  by assuming that there exist two uniform populations of genes and thus mass action kinetics is used to describe the activation process. We now show that the unnecessary underlying assumption of a uniform continuous population of genes in the cell can be overcome by using  the \emph{average dynamics} approach. In fact we  re-formulate the model by considering just two genes described by a MC on a finite state space.  

\subsubsection{The associated reaction scheme}
We keep the same notation as in \cite{VKBL} in order to simplify the comparison. So we use the following notation: 

\begin{itemize}
\item $A$ is the activator protein and $M_A$ its corresponding mRNA,
\item $R$ is the repressor protein and  $M_R$ its corresponding mRNA,
\item $C$ is a complex formed by $A$ and $R$.  
\end{itemize}

\noi Each gene can be either active or inactive, $D_A$, $D_R$ denote the inactive states, and $D'_A$, $D'_R$ the active states. We now collect all necessary reactions to model this situation:

\subsubsection*{Gene activation:}
$$\begin{array}{llll}
A+D_A\rlha[\theta_A]{\gamma_A} D'_{A},\\
A+D_R\rlha[\theta_R]{\gamma_R} D'_{R}\\
\end{array}$$

\subsubsection*{Transcription:}

$$\begin{array}{ll}
D_{A}\rightarrow^{\alpha_A} D_{A}+M_{A},~~~~~D'_{A}\rightarrow^{\alpha'_A} D'_{A}+M_{A}\\
D_{R}\rightarrow^{\alpha_R} D_{R}+M_{R},~~~~~D'_{R}\rightarrow^{\alpha'_R} D'_{R}+M_{R}\\
\end{array}$$

\subsubsection*{Translation:}
$$M_A\rightarrow^{\beta_A}A,~~~~~M_R\rightarrow^{\beta_R}R$$

\subsubsection*{Regulation and inhibition of protein $A$:}
$$A+R\rightarrow^{\gamma_C}C,~~~~~C\rightarrow^{\delta_A} R$$

\subsubsection*{Degradation:}

$$\begin{array}{ll}
M_A\rightarrow^{\delta_{MA}}\emptyset,~~~~~M_R\rightarrow^{\delta_{MR}}\emptyset,\\[4mm]
A\rightarrow^{\delta_{A}}\emptyset,~~~~~R\rightarrow^{\delta_{R}}\emptyset.
\end{array}$$

We next need to introduce the following concentrations for smaller molecules in the system and their complexes:

$$a=[A],~~c=[C],~~r=[R],~~ m_A=[M_A],~~ m_R=[M_R].$$

\subsubsection{A combined finite and infinite state space formulation}

\begin{figure}[htbp] 
   \centering
   \includegraphics[scale=0.4]{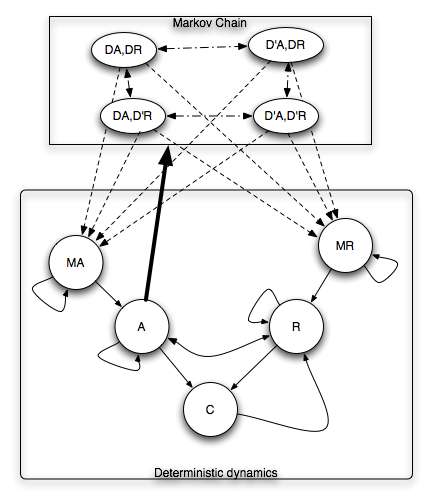} 
   \caption{Combined interaction graph $\vI_{C}$ for the VKBL model.}
   \label{fig:BL_cc_graph}
\end{figure}

To apply the averaging principle it is necessary to identify finite and infinite variables or degrees of freedom.  As discussed for genes interpreted as large molecular machines in perhaps single copy number the mass action kinetics typically cannot be used to describe the activation process.  Therefore it is a natural choice that the two genes $gene_A$ and $gene_R$ form the finite degrees of freedom whose time evolution will be given by a finite Markov chain. The concentrations of $A,C,R$ and $M_A, M_R$ will be naturally the continuous variables as these molecules are much smaller and very abundant relative to the genes. We first  describe the MC structure, and then state the equations for the concentrations $a,c,r$ derived by applying the principles of mass action kinetics. The structure of the dynamics can be represented through the combined \emph{interaction graph} in Figure \ref{fig:BL_cc_graph}. The structure of the links will be made clear by the construction of the MC and of the related deterministic dynamics.

\subsubsection{MC structure}
The Markov chain of the VKBL model has a state space given by all 
possible states of the couple $(gene_A,gene_R)$. 
$gene_A$ and $gene_R$ have state spaces respectively
$$\Sig_A=\{D_A,D'_A\},~~~\Sig_R=\{D_R,D'_R\}.$$ 
Therefore the total space is equal to

$$\Sig=\Sig_A\times\Sig_R=\{(D_A,D_R),(D'_A,D_R),(D_A,D'_R),(D'_A,D'_R)\}.$$

\noi The reactions associated to gene activation determine the infinitesimal rates of the MC. We consider transitions in which either $D_A$ or $D_R$ is affected as illustrated in Fig.
\ref{fig:BL_cc_graph}. Transitions like $(D_A,D_R)\ra (D'_A,D'_R)$ are excluded because they occur with a very small probability. In fact it is very unlikely that two molecules $A$ bind simultaneously on the two different existing binding sites. In order to appreciate the effects of describing $A$ protein with its concentration $a$ let us take the matrix $\Kd^T$ (the transpose of the infinitesimal generator) written in terms of the $A$ copy number always denoted by $a$:

\begin{equation}
\Kd^T\,=\left( \begin {array}{cccc} -a \left( \gamma_{{R}}+\gamma_{{A}}
 \right) &\theta_{{A}}&0&\theta_{{R}}\\\noalign{\medskip}\gamma_{{A}}a
&-\theta_{{A}}- \left( a-1 \right) \gamma_{{R}}&\theta_{{R}}&0
\\\noalign{\medskip}0& \left( a-1 \right) \gamma_{{R}}&-\theta_{{A}}-
\theta_{{R}}& \left( a-1 \right) \gamma_{{A}}\\\noalign{\medskip}a
\gamma_{{R}}&0&\theta_{{A}}&-\theta_{{R}}- \left( a-1 \right) \gamma_{
{A}}\end {array} \right).
\label{Kmatrix1}
\end{equation} 

\brem
The copy number $a$ is an integer greater or equal $1$. Note that we are not applying the average principle here in order to be able to discuss the transition of small particle to large particle copy numbers.
\erem

\noi The invariant measure is given by

{\small \begin{equation}
\begin{array}{ll}
\displaystyle\mu(a)\,=\,\left({\frac {\theta_{{A}}\theta_{{R}}}{\theta_{{A}}\theta_{{R}}+\theta_{{R
}}\gamma_{{A}}a+{a}^{2}\gamma_{{R}}\gamma_{{A}}-\gamma_{{R}}\gamma_{{A
}}a+\theta_{{A}}a\gamma_{{R}}}},{\frac {\theta_{{R}}\gamma_{{A}}a}{
\theta_{{A}}\theta_{{R}}+\theta_{{R}}\gamma_{{A}}a+{a}^{2}\gamma_{{R}}
\gamma_{{A}}-\gamma_{{R}}\gamma_{{A}}a+\theta_{{A}}a\gamma_{{R}}}},
\right.\\[5mm]
\displaystyle\left.{\frac {a\gamma_{{R}} \left( a-1 \right) \gamma_{{A}}}{\theta_{{A}}
\theta_{{R}}+\theta_{{R}}\gamma_{{A}}a+{a}^{2}\gamma_{{R}}\gamma_{{A}}
-\gamma_{{R}}\gamma_{{A}}a+\theta_{{A}}a\gamma_{{R}}}},{\frac {\theta_
{{A}}a\gamma_{{R}}}{\theta_{{A}}\theta_{{R}}+\theta_{{R}}\gamma_{{A}}a
+{a}^{2}\gamma_{{R}}\gamma_{{A}}-\gamma_{{R}}\gamma_{{A}}a+\theta_{{A}
}a\gamma_{{R}}}}
\right)
\end{array}
\label{mu_a}
\end{equation}}

\subsubsection{Large values of $A$ concentration and  independent genes}
Let us now consider the regime of large copy number of $A$, namely $a>>1$, this implies that $A$ is well described by concentration. In this case we could approximate $a-1\simeq a$. This would yield a MC with a new $\Kd^T$ given by:

\begin{equation}
{\Kd'}^T\,=\left( \begin {array}{cccc} -a \left( \gamma_{{R}}+\gamma_{{A}}
 \right) &\theta_{{A}}&0&\theta_{{R}}\\\noalign{\medskip}\gamma_{{A}}a
&-\theta_{{A}}- a \gamma_{{R}}&\theta_{{R}}&0
\\\noalign{\medskip}0& a \gamma_{{R}}&-\theta_{{A}}-
\theta_{{R}}& a\gamma_{{A}}\\\noalign{\medskip}a
\gamma_{{R}}&0&\theta_{{A}}&-\theta_{{R}}- a\gamma_{
{A}}\end {array} \right).
\label{Kmatrix2}
\end{equation} 
$\Kd'^T$ describes a product of two independent chains. In fact  
let us recall the structure of the spaces $\Sig_A=\{D_A,D'_A\},~~\Sig_R=\{D_R,D'_R\}.$ The matrix $\Kd'^T$ can be rewritten in the form (\ref{rates-for-product}) using the the following matrices

\begin{equation}
\Kd^T_A\,=\left( \begin {array}{cc} 
-a\gamma_A & \theta_A\\[3mm]
a\gamma_A & -\theta_A\\
\end {array} \right).
\label{MatrixA}
\end{equation} 

and

\begin{equation}
\Kd^T_R\,=\left( \begin {array}{cc} 
-a\gamma_R & \theta_R\\[3mm]
a\gamma_R & -\theta_R\\
\end {array} \right).
\label{MatrixR}
\end{equation} 

In the regime $a-1\simeq a$ the invariant measure is equal to

$$\begin{array}{ll}
\displaystyle \mu(a)\, = \,\left(\frac{\theta_A\,\theta_R}{(\theta_A+a\,\gamma_A)(\theta_R+a\,\gamma_R)},
\frac{a\,\gamma_A\,\theta_R}{(\theta_A+a\,\gamma_A)(\theta_R+a\,\gamma_R)},\right.\\[5mm]
\displaystyle\left.
\frac{a^2\,\gamma_A\,\gamma_R}{(\theta_A+a\,\gamma_A)(\theta_R+a\,\gamma_R)},
\frac{a\,\gamma_R\,\theta_A}{(\theta_A+a\,\gamma_A)(\theta_R+a\,\gamma_R)}\right).
\end{array}$$

\noi Now note that matrices $\Kd^T_A$ and $\Kd_B^T$ have invariant measures which are given by

$$\mu_A(a)=\left(\frac{\theta_A}{\theta_A+a\,\gamma_A},\frac{a\,\gamma_A}{\theta_A+a\,\gamma_A}\right)~\mbox{ 
and }~
\mu_R(a)=\left(\frac{\theta_R}{\theta_R+a\,\gamma_R},\frac{a\,\gamma_R}{\theta_R+a\,\gamma_R}\right).$$

\noi One can easily verify that the components of  the invariant measure $\mu(a)$ satisfy (\ref{prob}) and factor into the components of $\mu_A(a)$ and $\mu_R(a)$:

\begin{equation}
\mu(a)=(\mu_A^{(1)}(a)\,\mu_R^{(1)}(a)~,~\mu_A^{(2)}(a)\,\mu_R^{(1)}(a)~,~
\mu_A^{(2)}(a)\,\mu_R^{(2)}(a)~,~\mu_A^{(1)}(a)\,\mu_R^{(2)}(a)).
\label{mu_a_indip}
\end{equation}

\noi This corresponds to have a \emph{gene activation}  formed by the product of two independent MC's whose generators are respectively $\Kd_A$ and $\Kd_R$. Intuitively we can say that in the regime $a>>1$ 
the molecules activating the two genes are so abundant that genes do not compete and behave independently. The mechanism can be represented by the two reactions 

\begin{equation}
A+D_A\rlha[\theta_A]{\gamma_A} D'_{A},\mbox{ and }
A+D_R\rlha[\theta_R]{\gamma_R} D'_{R}.
\label{idip_reactions}
\end{equation}

\begin{remark}
Note that if $a>>1$ is not valid, then the dependence of the MC's can be interpreted as an cooperative effect, in the sense that the two MC's interact through the presence of the $A$ 
molecules. 
\end{remark}

The product of probabilities in (\ref{mu_a_indip}) indicates that the two MC's are independent whenever the number of $A$ molecules is so large that the reactions  (\ref{idip_reactions}) can be considered statistically independent.  In general one can envisage the possibility of having systems with large MC' s  interacting with a certain type of molecule. The chains do obviously compete for such molecules. As long as the number of molecules is small then the chains are in competition. But if the number of particles becomes large then the chains are effectively de-coupled.  The stochastic analysis and the related algebra of this problem will be developed in a different paper.

\subsubsection{Deterministic dynamics}
In the regime $a>>1$ the dynamics of $A,R$ and $C$ is obtained by using mass action kinetics. It turns out that the dynamical equations are  affected by  the MC only through the states
$M_A$ and $M_R$ (modelling the transcription process), and they are given by

\begin{equation}
\begin{array}{lll}
\dot{a}(t)=-\delta_A\,a(t)-\gamma_C\,a(t)\,r(t)+\beta_A\,m_A(t),\\[4mm]
\dot{c}(t)=-\delta_A\,c(t)+\gamma_C\,a(t)\,r(t),\\[4mm]
\dot{r}(t)=-\delta_R\,r(t)-\gamma_C\,a(t)\,r(t)+\beta_R\,m_R(t)+\delta_A\,c(t).
\end{array}
\label{cc_det_dyn}
\end{equation}

\noi Note that since the equation for $a(t), c(t)$ and $r(t)$ do not explicitly depend on 
the MC state, they are not affected by the averaging procedure.  Of course the dynamics  of the mRNA is dependent on the state of the genes. Let $f^{(\sig)}$ be the vector field corresponding to the state $\sig\in \Sig$. The transcription processes are governed by 
 
\begin{equation}
\begin{array}{lll}
\dot{m}_A(t)=f^{(D_A,.)}=\alpha_A-\delta_{MA}\,m_A(t)-\beta_A\,m_A(t),\\[4mm]
\dot{m}_A(t)=f^{(D'_A,.)}=\alpha'_A-\delta_{MA}\,m_A(t)-\beta_A\,m_A(t),\\[4mm]
\dot{m}_R(t)=f^{(.,D_R)}=\alpha_R-\delta_{MR}\,m_R(t)-\beta_R\,m_R(t),\\[4mm]
\dot{m}_R(t)=f^{(.,D'_R)}=\alpha'_R-\delta_{MR}\,m_R(t)-\beta_R\,m_R(t).
\end{array}
\label{cc_stoch_dyn}
\end{equation}

\subsection{Average dynamics for the VKBL model}
The average procedure does affect only $m_A$ and $m_R$. Indeed we find that

$$\dot{m}_A(t)=\sum_{D\in \{D_R,D'_R\}}\mu_{(D_A,D)}f^{(D_A,D)}+
\mu_{(D'_A,D)}f^{(D'_A,D)}$$

\noi and

$$\dot{m}_R(t)=\sum_{D\in \{D_A,D'_A\}}\mu_{(D,D_R)}f^{(D,D_R)}+
\mu_{(D,D'_R)}f^{(D,D'_R)}.$$

\noi After some algebra the average dynamics can be shown to be

\begin{equation}
\begin{array}{lllll}
\dot{a}(t)=-\delta_A\,a(t)-\gamma_C\,a(t)\,r(t)+\beta_A\,m_A(t), \\[4mm]
\dot{c}(t)=-\delta_A\,c(t)+\gamma_C\,a(t)\,r(t), \\[4mm]
\dot{r}(t)=-\delta_R\,r(t)-\gamma_C\,a(t)\,r(t)+\beta_R\,m_R(t)+\delta_A\,c(t), \\[5mm]
\displaystyle\dot{m}_{{A}}= 
{\frac {\alpha_{{A}}\theta_{{A}}}{\theta_{{A}}+\gamma_{{A}}a \left( t \right) }}+{\frac {\alpha'_{{{\it A}}}\gamma_{{A}}a \left( t \right) }{\theta_{{A}}+\gamma_{{A}}a \left( t \right) }}-\left( \delta_{{{\it MA}}}+\beta_{{A}} \right) m_{{A}} \left( t \right), \\[6mm]
\displaystyle \dot{m}_{{R}}= 
{\frac {\alpha_{{R}}\theta_{{R}}}{\theta_{{R}}+a \left( t \right) \gamma_{{R}}}}+{\frac {\alpha'_{{{\it R}}}a \left( t \right) \gamma_{{R}}}{\theta_{{R}}+a \left( t \right) \gamma_{{R}}}}
-\left( \delta_{{{\it MR}}}+\beta_{{R}} \right) m_{{R}} \left( t \right).
\end{array}
\label{cc_ave_dyn}
\end{equation}

The latter set of equations corresponds to the equations as stated in \cite{VKBL}, giving them a new interpretation.

 \begin{figure}[htbp] 
   \centering
   \includegraphics[scale=0.5]{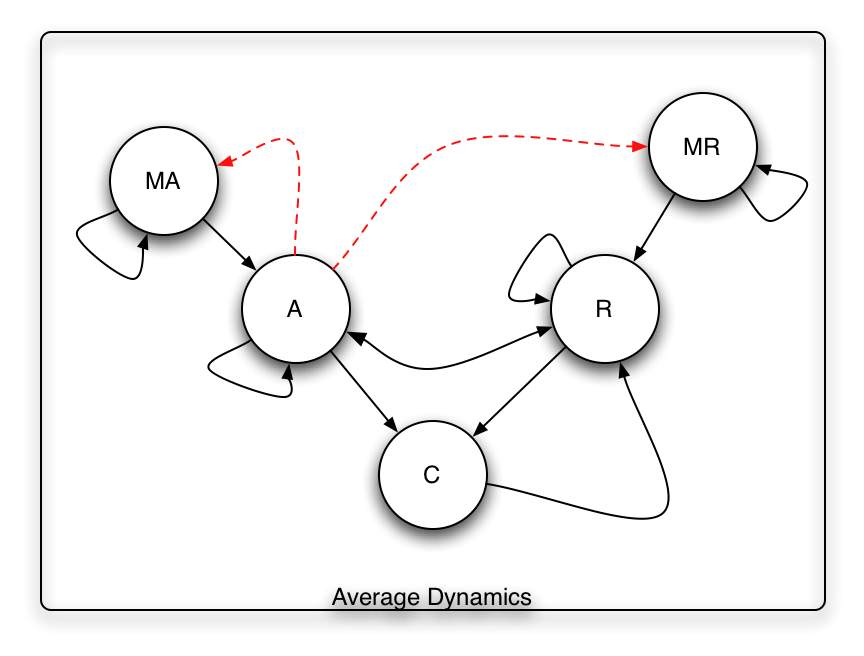} 
   \caption{The averaged interaction graph $\bar{\vI}$  for the VKBL model.}
   \label{fig:BL_cc_graph-average}
\end{figure}

 \noi By inspection of (\ref{cc_ave_dyn})  the averaged interaction graph $\bar{\vI}$ can be constructed 
 and becomes as shown in Fig.  \ref{fig:BL_cc_graph-average}. One can verify that new links are 
 present due to the effective transcription rates depending on $a$.

\subsubsection{Numerical analysis of the model} \label{subsub:oscillations}
The genetic oscillator described by the average dynamics can be investigated numerically. We use the values of parameters given in \cite{VKBL} and make the following choices:

$$\begin{array}{lll}
\alpha_A=50h^{-1},~~\alpha�_A=500 h^{-1},~~~\alpha_R=0.01 h^{-1},  \alpha'_R=50 h^{-1},~~\beta_A=50 h^{-1},~~\beta_R=5h^{-1},\\
\delta_{MA}=10 h^{-1},~~\delta_{MR}=0.5 h^{-1},~~\delta_{A}=1 h^{-1},~~\delta_R=0.2 h^{-1},\\
\gamma_{A}=1 h^{- 1}mol^{-1},~~\gamma_R=1 h^{-1}mol^{-1},~~\gamma_C=2h^{-1}mol^{-1},~~\theta_A=50 h^{-1},~~\theta_R=100 h^{-1}.
\end{array}$$

\begin{figure}[htbp] 
   \centering
   \includegraphics[scale=0.3]{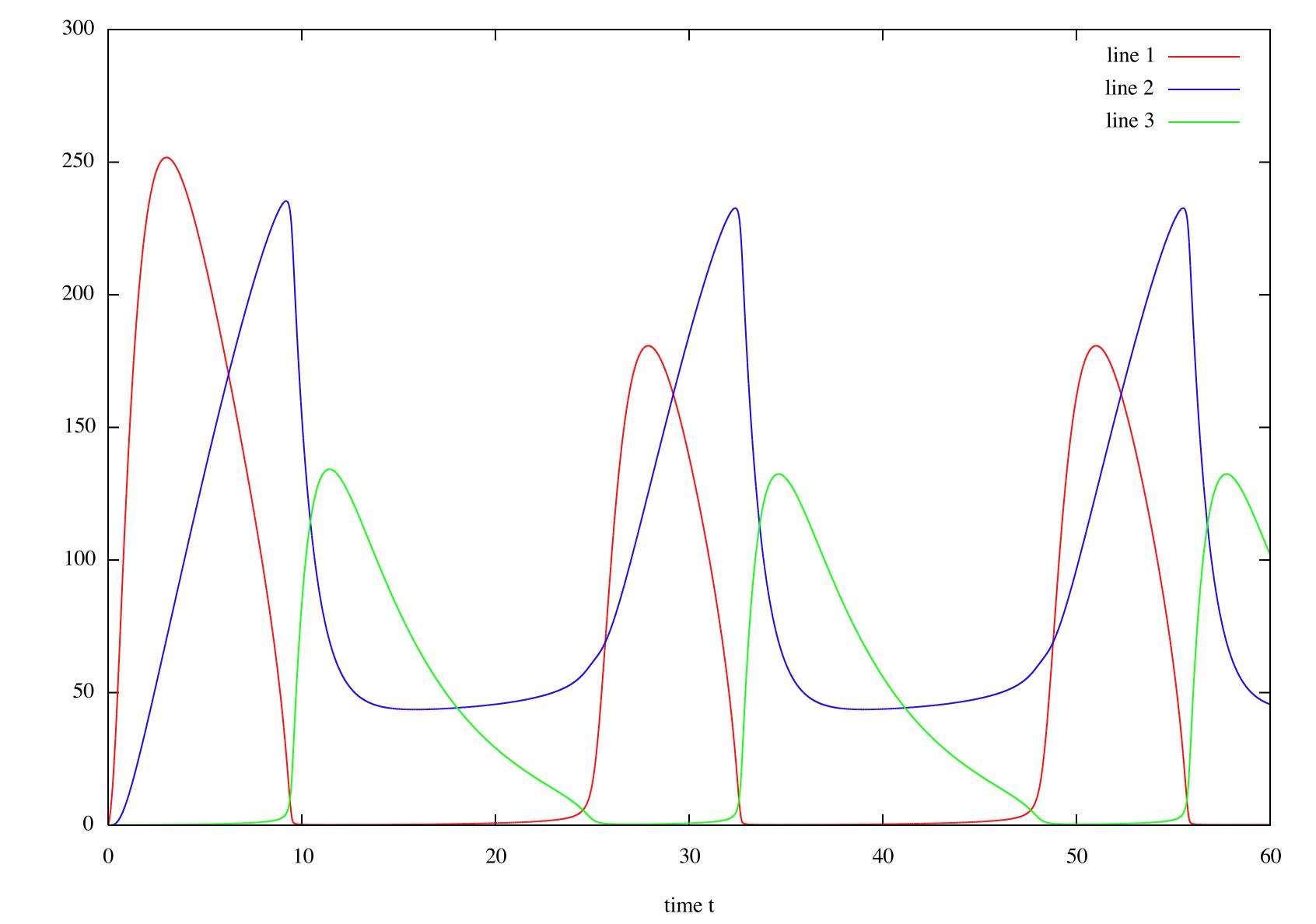} 
   \caption{A typical average dynamics  time evolution for proteins $A$ (red line 1), $C$ (blue line 2) and $R$ (green line 3) following a limit cycle.}
   \label{fig:ACRM}
\end{figure}  

\noi We integrate the average dynamics (\ref{cc_ave_dyn}) with an ODE solver numerically (using \textsc{GNU Octave}). Here we have chosen the initial conditions

$$a(0)=0,~~c(0)=0,~~r(0)=0,~~m_A(0)=0,~~m_R(0)=0.$$

\begin{figure}[htbp] 
   \centering
   \includegraphics[scale=0.3]{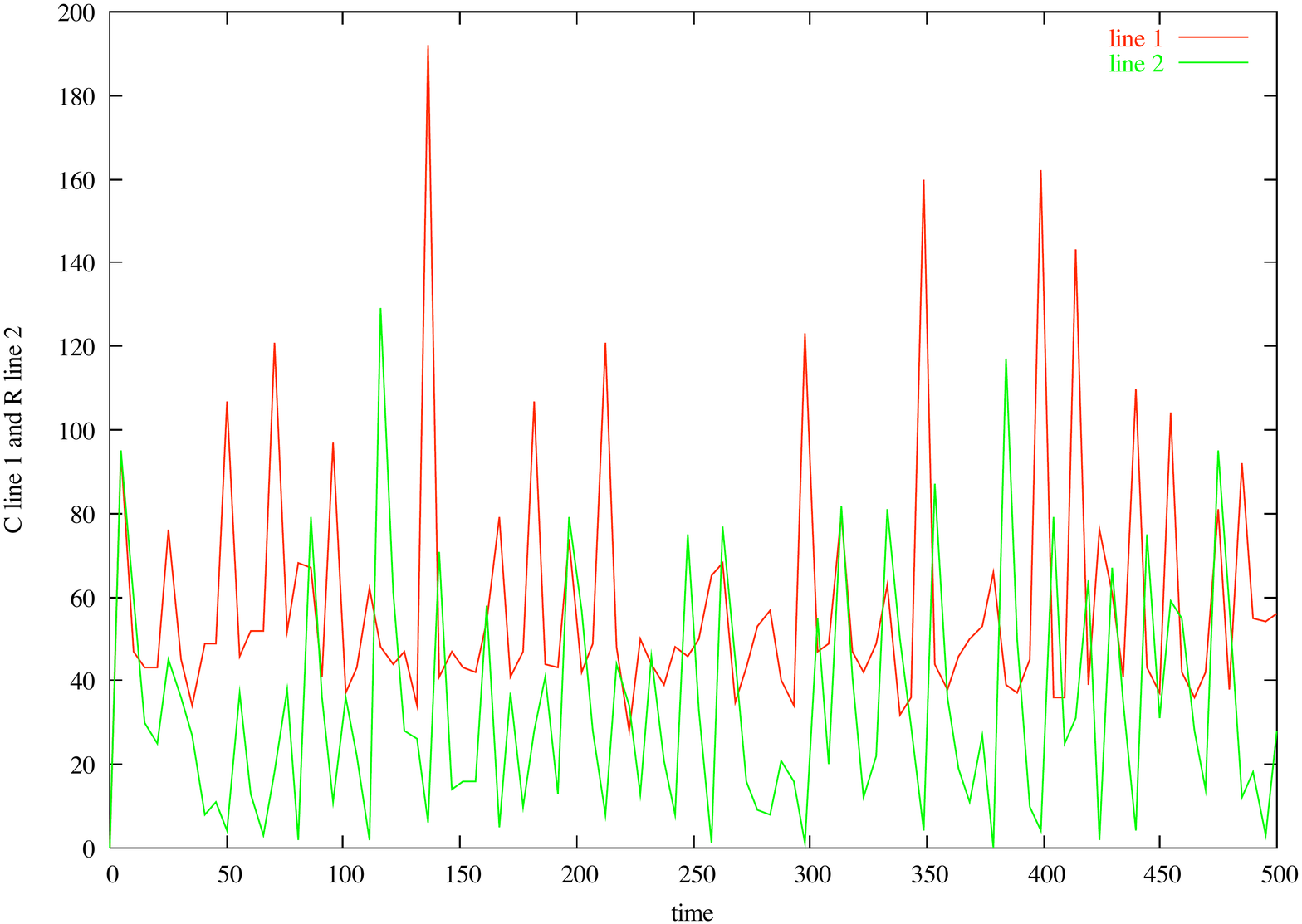} 
   \caption{A typical random time evolution of proteins $C$ and $R$ using Gillespie's algorithm in a stochastic version of the average dynamics (Randomness is due to finite particle numbers, not because of stochastic transitions in the original Markov chain part of the model).}
   \label{fig:CR_gillespie}
\end{figure}

\begin{figure}[htbp] 
   \centering
   \includegraphics[scale=0.3]{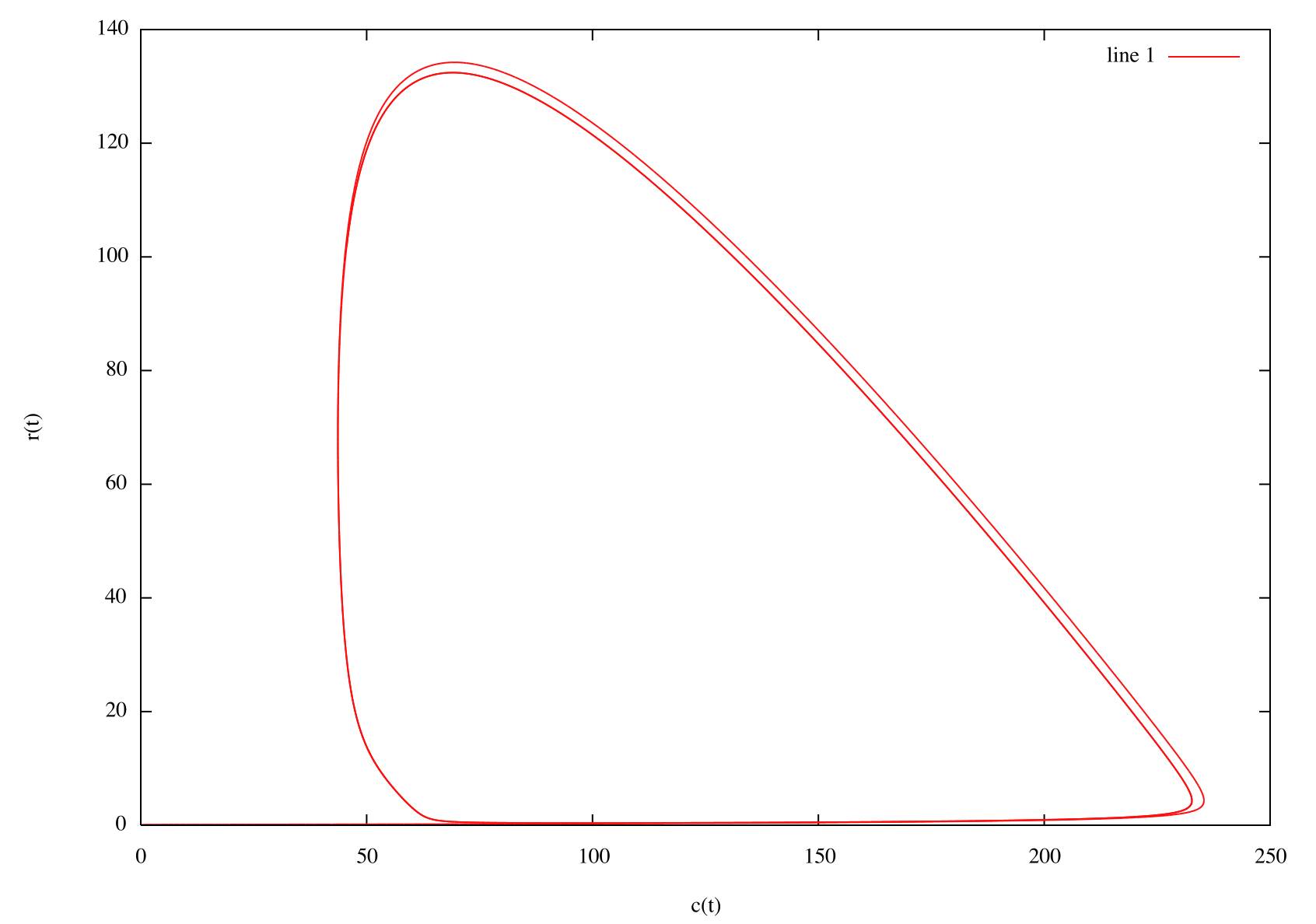} 
   \caption{Projection of the 'stochastic' limit cycle into the $c,r$ plane.}
   \label{fig:CR_limit_cycle}
\end{figure}

\noi In a second step we used \textsc{Dizzy} to simulate the average dynamics using Gillespie's algorithm, see Fig. \ref{fig:CR_gillespie}. This means we interpret the average dynamics vector field as chemical reactions where the particle number involved in the reaction stays finite. Similarly to the model presented in \cite{VKBL} we also found in this case numerically a  limit cycle. This is shown in Fig. \ref{fig:CR_limit_cycle}. In fact even though protein abundances now fluctuate in time this limit cycle is very close in average norm to the one created by the average dynamics. It holds of course that this similarity is higher if more particles are used in Gillespie's algorithm to simulate the dynamics. Note also that this limit cycle can be obtained analytically from the average dynamics directly when using the Hopf bifurcation theorem. One can verify that for the given parameter values there is one single unstable steady state which is located in the interior of the 'stochastic' limit cycle projected into the $c,r$-plane. 

\section{Conclusions}

In this paper we presented the average dynamics as a framework to derive  macroscopic models based on typical microscopic assumptions often occurring in complex systems theory. The system is separated into finite-state machines (themselves finite in number, typically occurring in a single copy number only), and system components that are associated with very abundant particles moving randomly and spatially homogeneously. These latter particles are given as concentrations, and they are typically describing particles communicating between the finite-state machines, creating switches between the states. The average dynamics is then naturally constructed out of a collection of deterministic vector fields and depending on a Markov chain steady state. The construction is a very useful tool to derive effective models that approximate stochastic systems described by  particular reaction schemes acting on a mesoscopic or microscopic scale. One of the best examples of such a situation is the use of combinations of Markov chains to describe conformational changes of large bio-molecules. In this case the average dynamics allows to establish a very modular approach to study complex bio-molecular  diagrams. Such models can be tested with data derived from measurements on different scales, and are therefore much better to test empirically. The framework described can also be interpreted as allowing to assign more or less automatically deterministic dynamics to given interaction diagrams, where the modeler can make choices whether a variable can have  a continuous range, or a discrete range, \cite{msgm-1,msgm-2}. The diagram then has a mathematical precise interpretation as an interaction graph associated to the dynamics. It can be further studied to investigate the qualitative behaviour of the system. The interaction graph is a more precise tool to identify feedback loops when compared to the motifs which as a static (i.e. time-independent) concept are sometimes inappropriately used in the same or similar role in Systems Biology \cite{review}. We made the step to use such graphs also in a multi-scale and hybrid dynamics setting.

\subsubsection*{Acknowledgements}
This paper is part of the research activities supported by 
UniNet contract 12990  funded by the 
European Commission in the context of the VI Framework Programme. We would like to thank all referees for very useful comments.

%


\begin{thebibliography}{999}


\bibitem{Amos}
M. Amos (Ed.): {\it Cellular computing}. SYSBIO, Oxford UP, 2004.

\bibitem{atkinson}
M.A. Atkinson et al.:  {\it Development of Genetic Circuitry Exhibiting Toggle Switch Or Oscillatory Behavior in scherichia Coli}. Cell, 113:5 (2003), 597-- 607.

\bibitem{Biggs}
N. Biggs: {\it Algebraic Graph Theory}. Cambridge University Press, 1993

\bibitem{hybrid1} M.S. Branicky: {\it Introduction to hybrid systems.} In D. Hristu-Varsakelis and W.S. Levine (eds.), Handbook of Networked and Embedded Control Systems,  Boston: Birkh\"auser, pp. 91-116, 2005. 

\bibitem{Domijan}
M. Domijan and M. Kirkilionis: {\it  Graph Theory and Qualitative Analysis of Reaction Networks}. Networks and Heterogeneous Media. 3:2 (2008), 295 -- 322.

\bibitem{elowitz}
M.B. Elowitz and S. Leibler: {\it A Synthetic Oscillatory Network of Transcriptional Regulators}. Nature, 403:6767 (2000), 335--338.

\bibitem{gardner}
T.S. Gardner, C.R. Cantor, J.J. Collins: {\it Construction of a Genetic Toggle Switch in Escherichia Coli}. Nature, 403:6767 (2000), 339--342.

\bibitem{Gibbons}
 A. Gibbons {\it Algorithmic Graph Theory}. Cambridge UP, 1985.
 
 \bibitem{Gillespie}
 D. T. Gillespie: {\it Stochastic simulation of chemical kinetics}. Annual Review of Physical Chemistry, 58 (2007), 35Ð-55.
 
\bibitem{Godsil}
C.D. Godsil and G. Royle: {\it Algebraic Graph Theory}. Springer Verlag, 2001.

\bibitem{2comp}
R. Guantes and J. F. Poyatos: {\it Dynamical Principles of Two-Component Genetic Oscillators}.
PLoS Comput Biol., 2:3 (2006), e30.

\bibitem{hasty}
J. Hasty, D. McMillen, J.J. Collins: {\it Engineered Gene Circuits}. Nature, 420:6912 (2002), 224--230.

\bibitem{julius}
A.A. Julius, M.S Sakar, A. Bemporad and G.J. Pappas: {\it  Hybrid model predictive control of induction of \emph{Escherichia coli} }. Proceedings of the 46th IEEE Conference on Decision and Control New Orleans, LA, USA, Dec. 12--14, 2007.

\bibitem{lemma}
J. Keilson: {\it Covariance and relaxation time in finite Markov chain}. Journal of Applied Mathematics and Stochastic Analysis, 11:3 (1998), 391--396.

\bibitem{msgm-1}
M. Kirkilionis, U. Janus, L. Sbano: {\it Multi-Scale Genetic Dynamic Modelling I : An Algorithm to Compute Generators. An Algorithmic
Markov Chain Based Approach}. WMI Preprint 4/2010, University of Warwick, 2010. go.warwick.ac.uk/maths

\bibitem{msgm-2}
M. Kirkilionis, U. Janus, L. Sbano: {\it Multi-Scale Genetic Dynamic Modelling II: Application to Synthetic Biology}. WMI Preprint 5/2010, University of Warwick, 2010. go.warwick.ac.uk/maths

\bibitem{LM-stability}
M. Kirkilionis and L. Sbano: {\it An Averaging Principle for Combined Interaction Graphs. Part II: Modularity and Perturbations of the Average Vector Field.}. In preparation.

\bibitem{review} M. Kirkilionis: Exploration of Cellular Reaction Systems. Briefings in Bioinformatics. 11:1 (2010), 153--178.

\bibitem{quasi} M. Kirkilionis and S. Walcher: {\it On comparison systems for ordinary differential equations}. Journal of Mathematical Analysis and Applications, 299:1 (2004), 157--173.

\bibitem{MA}
 S. Mangan and U. Alon: {\it Structure and function of the feed-forward loop network motif}. PNAS, 100:21 (2003), 11980--11985.

\bibitem{hybrid}
X. Mao and C. Yuan: {\it Stochastic Differential Equations With Markovian Switching}.  Imperial College Press, 2006.

\bibitem{MM}
M. Marcus and H. Minc: {\it A survey of matrix theory and matrix inequalities}. Prindle, Weber and Schmidt, Boston, 1996.

\bibitem{precision}
L.G. Morelli and F. J\"ulicher: {\it Precision of Genetic Oscillators and Clocks}. Phys Rev Lett., 98:22 (2007), 228101.

\bibitem{Pinsky}
M. Pinsky: {\it Differential Equations with a Small Parameter and the Central. Limit Theorem for Functions Defined on a Finite Markov chain}.  Z. Wahrscheinlichkeitstheorie und verw. Geb., 9 (1968), 101--111.

\bibitem{LM1}
L. Sbano and M. Kirkilionis: {\it Molecular Systems with Infinite and Finite Degrees of Freedom. Part I: Multi-Scale analysis}. WMI Preprint 5/2007. Available at: arXiv:0802.4259v2 [q-bio.BM], 2008.

\bibitem{LM2}
L. Sbano and M. Kirkilionis: {\it Molecular Systems with Infinite and Finite Degrees of Freedom. Part II: Deterministic Dynamics and Examples}. WMI Preprint 7/2007. Available at: arXiv:0802.4279v1 [q-bio.MN], 2008.

\bibitem{LM3}
L. Sbano and M. Kirkilionis: {\it Multiscale analysis of reaction networks}. 
Theory Biosciences 127 (2008), 107--123.

\bibitem{smith}  H.L. Smith: Monotone Dynamical Systems: An Introduction to the Theory of Competitive and Cooperative Systems. American Mathematical Society, Providence, Rhode Island, 1995.

\bibitem{Stroock}
D.W. Stroock: {\it An introduction to Markov Processes}. Graduate Texts, Springer Verlag, 2005.

\bibitem{VKBL}
J.M.G. Vilar, H.Y. Kueh, N. Barkai, and S. Leibler: {\it Mechanisms of noise-resistance in genetic oscillators}.
 PNAS  9:99 (2002),  5988--5992.
 
 \bibitem{WC}
G.G. Walter and M. Contraras: {\it Compartmental Modelling with Networks}. Bikh\"auser, 1999.

\end{thebibliography}
\end{document}